\documentclass[aps,onecolumn,10pt]{revtex4}
\usepackage{amsmath}
\usepackage{amssymb}
\usepackage{hyperref}
\hypersetup{colorlinks=true,linkcolor=red,citecolor=green,urlcolor=blue}
\numberwithin{equation}{section}
\numberwithin{equation}{section}

\begin{document}
\allowdisplaybreaks
\setcounter{equation}{0}

\title{Properties of associated Legendre conical functions }

\author{Daniel A. Norman, Philip D. Mannheim and Tianye Liu }
\affiliation{Department of Physics, University of Connecticut, Storrs, CT 06269, USA \\
daniel.norman@uconn.edu, philip.mannheim@uconn.edu, tianye.liu@uconn.edu\\ }

\date{August 1  2025}

\begin{abstract}
We present some new properties of associated Legendre conical functions of the first and second kind, $P^{-1/2-K}_{-1/2+i \tau}(\chi)$ and $Q^{-1/2-K}_{-1/2+i \tau}(\chi)$. In particular we show that with the $\tau$-independent  $\mathcal{R}^{K}_{n}(\chi)=(2\pi)^{-3/2}\tanh^{-K}\chi\sinh^{-1/2}\chi[\Gamma(1+K)]^{-1}\int_{0}^{2\pi} d\omega \left (1-\cos \omega/\cosh\chi\right)^{K}e^{in\omega}$ for any general $K$,  we can set $P^{-1/2-K}_{-1/2+i \tau}(\chi)=2\sum_{n}\mathcal{R}^{K}_{n}(\chi)\sin[(\tau-in)\chi)]/(\tau-in)$, where $n$ ranges from $-\ell$ to $\ell$ in unit steps when $K$ is a non-negative integer $\ell$, and from $-\infty$ to $\infty$ in unit steps otherwise. Also we can set $Q^{-1/2-K}_{-1/2+i \tau}(\chi)=-\pi\sum_n\mathcal{R}^K_{n-K}(\chi)e^{-i \chi(\tau-i(n-K))}/(\tau-i(n-K))$, where $n$ ranges from $0$ to $2\ell$ in unit steps when $K$ is a non-negative integer $\ell$, and from $0$ to $\infty$ in unit steps otherwise. With these forms isolating the entire $\tau$ dependence, and especially its associated pole structure,  we can use these forms to determine  closed form expressions for integrals over $\tau$ of  associated Legendre conical functions and their products. The $Q^{-1/2-K}_{-1/2+i \tau}(\chi)$ have an integral  representation containing the integral $\int_{\chi}^{\infty}d\omega e^{i\omega\tau}(\cosh\omega-\cosh\chi)^{K}$, an integral that only converges at $\omega=\infty$ if ${\rm Re}[K]<{\rm Im}[\tau]$. We show how to use the divergence of this integral outside of this range in order to characterize the complex $\tau$ plane pole structure of  $Q^{-1/2-K}_{-1/2+i \tau}(\chi)$. We present a  new treatment of the Borwein integral and the Nyquist-Shannon sampling theorem.

\end{abstract}

\maketitle

\section{Introduction}
\label{S1}

The associated Legendre functions are solutions to the second-order differential equation (see e.g. \cite{AS,GR,digital})
\begin{align}
\left[(z^2-1)\frac{d^2}{dz^2}+2z\frac{d}{dz}-\nu(\nu+1)-\frac{\mu^2}{z^2-1}\right]F(\nu,\mu,z)=0,
\label{1.1}
\end{align}
an equation that when $z>1$ can be written in the form
\begin{align}
\left[\frac{d^2}{d\chi^2}+\frac{\cosh\chi}{\sinh\chi}\frac{d}{d\chi}-\nu(\nu+1)-\frac{\mu^2}{\sinh^2\chi}\right]F(\nu,\mu,\chi)=0,
\label{1.2}
\end{align}
 where $z=\cosh\chi$. As such, associated Legendre functions generalize the standard associated Legendre polynomials  to non-integer $\nu$ and $\mu$ \cite{footnote1}. 
 With (\ref{1.2}) being a second-order differential equation, there are two classes of solution: $P^{\mu}_{\nu}(\chi)$ and $Q^{\mu}_{\nu}(\chi)$, solutions that   are respectively known as being of the first and second kind. As given in \cite{AS,GR,digital}, for the real, positive $\chi$ of interest to us here  they can be represented by hypergeometric functions 
\begin{align}
P^{\mu}_{\nu}(\chi)&=\frac{1}{ \Gamma(1-\mu)}\left(\frac{\cosh\chi+1}{\cosh\chi-1}\right)^{\mu/2}F(-\nu,\nu+1;1-\mu;(1-\cosh\chi)/2),
\nonumber\\
Q^{\mu}_{\nu}(\chi)&=\frac{e^{i\mu\pi}\pi^{1/2}\Gamma(\nu+\mu+1)}{ 2^{\nu+1}\Gamma(\nu+3/2)}\frac{(\sinh\chi)^{\mu}}{(\cosh\chi)^{\nu+\mu+1}}
F(\nu/2+\mu/2+1,\nu/2+\mu/2+1/2;\nu+3/2;1/\cosh^2\chi),
\label{1.3}
\end{align}
or by integrals 
\begin{align}
P^{\mu}_{\nu}(\chi)=&\left(\frac{2}{\pi \sinh\chi}\right)^{1/2}\frac{\tanh^{\mu+1/2}\chi}{\Gamma(1/2-\mu)}\int_0^{\chi}d\omega \cosh((\nu+1/2)\omega)\left(1-\frac{\cosh\omega}{\cosh\chi}\right)^{-\mu-1/2},\qquad [{\rm Re}[\mu]<1/2],
\nonumber\\
Q^{\mu}_{\nu}(\chi)=&\left(\frac{\pi}{2 \sinh\chi}\right)^{1/2}\frac{e^{i\pi\mu}\tanh^{\mu+1/2}\chi}{\Gamma(1/2-\mu)}\int_{\chi}^{\infty}d\omega e^{-(\nu+1/2)\omega}\left(\frac{\cosh\omega}{\cosh\chi}-1\right)^{-\mu-1/2}, \qquad [{\rm Re}[\mu]<1/2,~{\rm Re}[\nu+\mu]>-1],
\label{1.4}
\end{align}
 with the required parameter ranges for the integrals to exist being indicated.
 
The conical functions are special cases of these functions with $\nu=-1/2+i\tau$ where $\tau$ is real, a condition under which $P^{\mu}_{-1/2+i\tau}(\chi)=P^{\mu}_{-1/2-i\tau}(\chi)$, though  $Q^{\mu}_{-1/2+i\tau}(\chi)\neq Q^{\mu}_{-1/2-i\tau}(\chi)$. (While $\tau$ is taken to be real in order to define conical functions we note that we shall have occasion to be interested in complex $\tau$ in the following.) And with $\mu=-1/2-K$ with arbitrary $K$ the conical functions obey   
\begin{align}
\left[\frac{d^2}{d\chi^2}+\frac{\cosh\chi}{\sinh\chi}\frac{d}{d\chi}+\tau^2+\frac{1}{4}-\frac{(-1/2-K)^2}{\sinh^2\chi}\right]F(\tau,K,\chi)=0,
\label{1.5}
\end{align}
and have integral representations of the form
\begin{align}
P^{-1/2-K}_{-1/2+i\tau}(\chi)&=\left(\frac{2}{\pi \sinh\chi}\right)^{1/2}\frac{\tanh^{-K}\chi}{\Gamma(1+K)}\int_0^{\chi}d\omega \cos(\omega\tau)\left(1-\frac{\cosh\omega}{\cosh\chi}\right)^{K}
\nonumber\\
&=\left(\frac{1}{2\pi \sinh\chi}\right)^{1/2}\frac{\tanh^{-K}\chi}{\Gamma(1+K)}\int_{-\chi}^{\chi}d\omega e^{-i\omega\tau}\left(1-\frac{\cosh\omega}{\cosh\chi}\right)^{K}, \qquad [{\rm Re}[K]>-1],
\nonumber\\
Q^{-1/2-K}_{-1/2+i\tau}(\chi)&=-\left(\frac{\pi}{2 \sinh\chi}\right)^{1/2}\frac{ie^{-i\pi K}\tanh^{-K}\chi}{\Gamma(1+K)}\int_{\chi}^{\infty}d\omega e^{-i\omega\tau}\left(\frac{\cosh\omega}{\cosh\chi}-1\right)^{K},\qquad [-1<{\rm Re}[K]<-{\rm Im}[\tau]],
\nonumber\\
Q^{-1/2-K}_{-1/2-i\tau}(\chi)&=-\left(\frac{\pi}{2 \sinh\chi}\right)^{1/2}\frac{ie^{-i\pi K}\tanh^{-K}\chi}{\Gamma(1+K)}\int_{\chi}^{\infty}d\omega e^{i\omega\tau}\left(\frac{\cosh\omega}{\cosh\chi}-1\right)^{K},\qquad [-1<{\rm Re}[K]<{\rm Im}[\tau]],
\label{1.6}
\end{align}
with the required parameter ranges for the integrals to exist  again being indicated. When both $\tau$ and $K$ are real the solutions to (\ref{1.5}) are either real ($P^{-1/2-K}_{-1/2+i\tau}(\chi)$), or up to a common overall factor  in complex conjugate pairs ($Q^{-1/2-K}_{-1/2+i\tau}(\chi)$ and $Q^{-1/2-K}_{-1/2-i\tau}(\chi))$.

The conical functions arise in a variety of situations, and with $\mu=-1/2-\ell$ with integer, non-negative $\ell$  and real $\tau$  the $Y^m_{\ell}(\theta,\phi)P^{-1/2-\ell}_{-1/2+i\tau}(\chi)/\sinh^{1/2}\chi$  are the wave functions of  scalar fields propagating in a space of constant  3-curvature $k$ that is negative with line element $ds^2=(-k)^{-1}[d\chi^2+\sinh^2\chi(d\theta^2+\sin^2\theta d\phi^2)]$. With $\tau$ being taken to be real and continuous between $0$ and $\infty$ these  modes are then complete.  Spaces of constant negative 3-curvature are of interest to cosmological theory, and especially in the conformal gravity theory as  described for instance in \cite{Mannheim2006,Mannheim2017}. In addition,  in this same theory the temporal behavior of modes fluctuating around a background space of constant negative curvature are  also described by conical functions with  $\nu=-1/2+i\tau$, but with $\mu=-1/2-K$, where $K$ is not required to be an integer \cite{Mannheim2020,Amarasinghe2021b}.

For some specific values of $\tau$ and $K$ the conical functions  can be expressed in terms of elementary functions. Thus for $\mu=1/2$, $\mu=-1/2$ and $\mu=-3/2$ and arbitrary $\tau$ for instance  we have 
\begin{align}
P^{1/2}_{-1/2+\pm i\tau}(\chi)&= \left(\frac{2}{\pi\sinh\chi}\right)^{1/2}\cos(\tau\chi),\qquad P^{-1/2}_{-1/2+\pm i\tau}(\chi)= \left(\frac{2}{\pi\sinh\chi}\right)^{1/2}\frac{\sin(\tau\chi)}{\tau},
\nonumber\\
P^{-3/2}_{-1/2\pm i\tau}(\chi)&= \frac{1}{(\tau^2+1)}\left(\frac{2}{\pi\sinh\chi}\right)^{1/2}\left[\frac{\cosh\chi}{\sinh\chi}\frac{\sin(\tau\chi)}{\tau}-\cos(\tau\chi)\right],
\nonumber\\
Q^{1/2}_{-1/2+ i\tau}(\chi)&=i \left(\frac{\pi}{2\sinh\chi}\right)^{1/2}e^{-i\tau\chi},\qquad Q^{1/2}_{-1/2-i\tau}(\chi)= i\left(\frac{\pi}{2\sinh\chi}\right)^{1/2}e^{i\tau\chi},
\nonumber\\
Q^{-1/2}_{-1/2+i\tau}(\chi)&=- \left(\frac{\pi}{2\sinh\chi}\right)^{1/2}\frac{e^{-i\tau\chi}}{\tau},\qquad Q^{-1/2}_{-1/2-i\tau}(\chi)= \left(\frac{\pi}{2\sinh\chi}\right)^{1/2}\frac{e^{i\tau\chi}}{\tau},
&\nonumber\\
Q^{-3/2}_{-1/2+i\tau}(\chi)&= \frac{1}{(\tau^2+1)}\left(\frac{\pi}{2\sinh\chi}\right)^{1/2}\left[-\frac{\cosh\chi}{\sinh\chi}\frac{e^{-i\tau\chi}}{\tau}-ie^{-i\tau\chi}\right],
\nonumber\\
Q^{-3/2}_{-1/2-i\tau}(\chi)&= \frac{1}{(\tau^2+1)}\left(\frac{\pi}{2\sinh\chi}\right)^{1/2}\left[\frac{\cosh\chi}{\sinh\chi}\frac{e^{i\tau\chi}}{\tau}-ie^{i\tau\chi}\right].
\label{1.7}
\end{align}
The integrals in (\ref{1.6}) immediately recover the expressions for  $P^{-1/2}_{-1/2+ i\tau}(\chi)$ and $P^{-3/2}_{-1/2- i\tau}(\chi)$. We recover the expressions for $Q^{-1/2}_{-1/2\pm i\tau}(\chi)$ and $Q^{-3/2}_{-1/2+\pm  i\tau}(\chi)$ from the integrals in (\ref{1.6})  if we respectively replace $\tau$ by $\tau-i\epsilon$ and $\tau+i\epsilon$  (i.e., we consider $Q^{-1/2-\ell}_{-1/2+ i(\tau-i\epsilon)}(\chi)$ and $Q^{-1/2-\ell}_{-1/2- i(\tau+i\epsilon)}(\chi)$ for $\ell=(0,1)$) so as to get convergence at $\omega=\infty$. For the $P^{1/2}_{-1/2\pm i\tau}(\chi)$ and $Q^{1/2}_{-1/2+\pm i\tau}(\chi)$ we use the connection formulae \cite{AS,GR,digital}
\begin{align}
&P^{1/2+K}_{-1/2\pm i\tau}(\chi)=\frac{\Gamma(\pm i\tau+K+1)}{\Gamma(\pm i\tau-K)}\left[P^{-1/2-K}_{-1/2\pm i\tau}(\chi)+\frac{2i}{\pi}e^{i\pi K}\cos(K\pi)Q^{-1/2-K}_{-1/2\pm i\tau}(\chi)\right],
\nonumber\\
&Q^{1/2+K}_{-1/2\pm i\tau}(\chi)=-e^{2i\pi K}\frac{\Gamma(\pm i\tau+K+1)}{\Gamma(\pm i\tau-K)}Q^{-1/2-K}_{-1/2\pm i\tau}(\chi).
\label{1.8}
\end{align}

The typical $P^{-1/2}_{-1/2+i\tau}(\chi)$ function is well behaved at $\chi=0$, while the typical  $Q^{-1/2}_{-1/2+i\tau}(\chi)$ function is singular  at $\chi=0$ (analogous to the regular and irregular Bessel functions, to which the conical functions  can be related in the large $\tau$ limit \cite{Cohl2018}). Additionally, the function $P^{-1/2}_{-1/2+i\tau}(\chi)$ is well behaved at $\tau=0$, while the function $Q^{-1/2}_{-1/2+i\tau}(\chi)$ is singular  at $\tau=0$.  With the  range of the  integral representation of the general $P^{-1/2-K}_{-1/2\pm i\tau}(\chi)$ being finite, the $P^{-1/2-K}_{-1/2\pm i\tau}(\chi)$ cannot have any complex $\tau$ plane poles.  With the  range of the  integral representation of the $Q^{-1/2-K}_{-1/2\pm i\tau}(\chi)$ being infinite, the $Q^{-1/2-K}_{-1/2\pm i\tau}(\chi)$ not only can but in fact do have complex $\tau$ plane poles.  We will explore the explicit complex $\tau$ plane structure of both the $P^{-1/2-K}_{-1/2\pm i\tau}(\chi)$ and $Q^{-1/2-K}_{-1/2\pm i\tau}(\chi)$ functions in the following.

For general integer non-negative $\ell$ the conical functions can also be written in a closed form  as \cite{Bander1966,Mannheim2020}  
\begin{align}
P^{-1/2-\ell}_{-1/2+i\tau}(\chi)&=(-1)^{\ell+1}\left(\frac{\pi}{2}\right)^{1/2}N^2_{\ell}(\tau)\sinh^{1/2}\chi\sinh^{\ell}\chi\left( \frac{1}{\sinh\chi}\frac{d}{d\chi}\right)^{\ell+1}\cos(\tau\chi),
\label{1.9}
\end{align}
and \cite{Mannheim2020}
\begin{align}
Q^{-1/2-\ell}_{-1/2+i\tau}(\chi)&=i(-1)^{\ell+1}\left(\frac{\pi}{2}\right)^{3/2} N^2_{\ell}(\tau)\sinh^{1/2}\chi\sinh^{\ell}\chi\left( \frac{1}{\sinh\chi}\frac{d}{d\chi}\right)^{\ell+1}e^{-i\tau\chi},
\label{1.10}
\end{align}
where \cite{footnote2}
\begin{align}
N_{\ell}(\tau)=\left(\frac{2}{\pi\tau^2(\tau^2+1^2)....(\tau^2+\ell^2)}\right)^{1/2}=\left(\frac{2}{\tau\sinh(\pi\tau)\Gamma(\ell+1+i\tau)\Gamma(\ell+1-i\tau)}\right)^{1/2}.
\label{1.11}
\end{align}
The reason for the $N_{\ell}(\tau)$ factor is twofold. First, this is the normalization needed so that (\ref{1.9}) and (\ref{1.10}) are normalized to the hypergeometric functions as  given in (\ref{1.3}). Under this normalization $P^{-1/2-\ell}_{-1/2+i\tau}(\chi)$ and $Q^{-1/2-\ell}_{-1/2+i\tau}(\chi)$ as given in (\ref{1.9}) and (\ref{1.10}) obey the recurrence relations \cite{AS,GR,digital}
\begin{align}
&[\tau^2+(\ell+1)^2]P^{-3/2-\ell}_{-1/2+i\tau}(\chi)=(2\ell+1)\frac{\cosh\chi}{\sinh\chi}P^{-1/2-\ell}_{-1/2+i\tau}(\chi)-P^{1/2-\ell}_{-1/2+i\tau}(\chi),
\nonumber\\
&[\tau^2+(\ell+1)^2]Q^{-3/2-\ell}_{-1/2+i\tau}(\chi)=(2\ell+1)\frac{\cosh\chi}{\sinh\chi}Q^{-1/2-\ell}_{-1/2+i\tau}(\chi)-Q^{1/2-\ell}_{-1/2+i\tau}(\chi),
\label{1.12}
\end{align}
so that starting with the values given in (\ref{1.7}) (values that (\ref{1.9}) and (\ref{1.10}) immediately recover for $P^{-1/2}_{-1/2+i\tau}(\chi)$ and $Q^{-1/2}_{-1/2+i\tau}(\chi)$), we can then generate $P^{-1/2-\ell}_{-1/2+i\tau}(\chi)$ and $Q^{-1/2-\ell}_{-1/2+i\tau}(\chi)$ for arbitrary positive, integer $\ell$, to precisely give the general $\ell$ expressions given in (\ref{1.9}) and  (\ref{1.10}) \cite{footnote3}. Secondly, the $N_{\ell}(\tau)$ factor serves as a normalization factor for the $P^{-1/2-\ell}_{-1/2+i\tau}(\chi)$ according to the normalization integral \cite{Bander1966}
\begin{align}
&\int_0^{2\pi}d\phi
\int_0^{\pi} d\theta \sin\theta Y_{\ell}^{*m}(\theta,\phi)Y^{m^{\prime}}_{\ell^{\prime}}(\theta,\phi)\int_0^{\infty}d\chi \sinh^2\chi \gamma^*_{\tau,\ell}(\chi)\gamma_{\tau^{\prime},\ell^{\prime}}(\chi)
\nonumber\\
&=\delta_{\ell,\ell^{\prime}}\delta_{m,m^{\prime}}\int_0^{\infty}d\chi \sinh^2\chi \gamma^*_{\tau,\ell}(\chi)\gamma_{\tau^{\prime},\ell^{\prime}}(\chi)=\delta(\tau-\tau^{\prime})\delta_{\ell,\ell^{\prime}}\delta_{m,m^{\prime}}, 
\label{1.13}
\end{align}
and  the addition theorem \cite{Bander1966}
\begin{align}
&\sum^{\infty}_{\ell=0}\sum_{m=-\ell}^{\ell}\gamma^*_{\tau,\ell}(\chi_1)\gamma_{\tau,\ell}(\chi_2)Y^{*m}_{\ell}(\theta_1,\phi_1)Y^{m}_{\ell}(\theta_2,\phi_2)
=\sum^{\infty}_{\ell=0}\gamma^*_{\tau,\ell}(\chi_1)\gamma_{\tau,\ell}(\chi_2)\frac{(2\ell+1)}{4\pi}P_{\ell}(\cos\Theta)=\frac{\tau \sin(\tau\psi)}{2\pi^2\sinh\psi},
\label{1.14}
\end{align}
where
\begin{align}
\gamma_{\tau,\ell}(\chi)&=(-1)^{\ell+1}\left(\frac{2}{\pi\sinh\chi}\right)^{1/2}\frac{P^{-1/2-\ell}_{-1/2+i\tau}(\chi)}{N_{\ell}(\tau)}
=(-1)^{\ell+1}[\tau^2(\tau^2+1)....(\tau^2+\ell^2)]^{1/2}\frac{P^{-\ell-1/2}_{-1/2+i\tau}(\chi)}{\sinh^{1/2}\chi},
\nonumber\\
 \cos\Theta&=\cos\theta_1\cos\theta_2+\sin(\phi_1-\phi_2)\sin\theta_1\sin\theta_2, \qquad
\cosh\psi=\cosh\chi_1\cosh\chi_2-\cos\Theta\sinh\chi_1\sinh\chi_2.
\label{1.15}
\end{align}
There is no analogous $\chi$ normalization integral for the $Q^{-1/2-\ell}_{-1/2+i\tau}(\chi)$ as they are not normalizable  (they diverge at $\chi=0$).  Thus we cannot introduce the $N_{\ell}(\tau)$ factor or fix the overall normalization for the   $Q^{-1/2-\ell}_{-1/2+i\tau}(\chi)$ from properties involving the $Q^{-1/2-\ell}_{-1/2+i\tau}(\chi)$ sector alone. We can, however,  instead fix the $Q^{-1/2-\ell}_{-1/2+i\tau}(\chi)$ normalization factors from the general connection formulae  \cite{AS,GR,digital}
\begin{align}
i\pi e^{-iK\pi}\sin(i\tau\pi)P^{-1/2-K}_{-1/2+i\tau}(\chi)&=\sin((i\tau-K)\pi)Q^{-1/2-K}_{-1/2+i\tau}(\chi)+\sin((i\tau+K)\pi)Q^{-1/2-K}_{-1/2-i\tau}(\chi),
\nonumber\\
(-1)^K\pi \sinh(\tau\pi) P^{-1/2-K}_{-1/2+i\tau}(\chi)=&-i\sinh(\tau\pi)\cos(K\pi)[Q^{-1/2-K}_{-1/2+i\tau}(\chi)+Q^{-1/2-K}_{-1/2-i\tau}(\chi)]
\nonumber\\
&+\cosh(\tau\pi)\sin(K\pi)[Q^{-1/2-K}_{-1/2+i\tau}(\chi)-Q^{-1/2-K}_{-1/2-i\tau}(\chi)],
\label{1.16}
\end{align}
\begin{align}
Q^{-1/2-K}_{-1/2+i\tau}(\chi)&=\frac{i\pi e^{-iK\pi}}{2\cos(K\pi)}\left[P^{-1/2-K}_{-1/2+i\tau}(\chi)-\frac{\Gamma(i\tau-K)}{\Gamma(i\tau+1+K)}P^{1/2+K}_{-1/2+i\tau}(\chi)\right]
\label{1.17}
\end{align}
that relate the $Q^{-1/2-\ell}_{-1/2+i\tau}(\chi)$ and the $P^{-1/2-\ell}_{-1/2+i\tau}(\chi)$ as we have fixed the normalization of the $P^{-1/2-\ell}_{-1/2+i\tau}(\chi)$.
As constructed, the
$P^{-1/2-\ell}_{-1/2+i\tau}(\chi)$ function is well behaved at $\chi=0$ and pole free, while the function $Q^{-1/2-\ell}_{-1/2+i\tau}(\chi)$ is singular  at $\chi=0$ and has poles in the complex $\tau$ plane. An analogous behavior is met for the general conical functions with $\mu=-1/2-K$ where $K$ is arbitrary. Specifically, while the $P^{-1/2-K}_{-1/2+i\tau}(\chi)$ remain singularity free, for any non-integer but real $K$ the $Q^{-1/2-K}_{-1/2+i\tau}(\chi)$  have,  as discussed in \cite{Liu2024} and below, a very intricate singularity structure in the complex $\tau$ plane. 

Unlike the $K=\ell$ modes the general $K$ modes cannot immediately be written in a closed form. It is the purpose of this paper to provide closed forms in the general $K$ case, while also providing closed forms for $\tau$ integrals such as $\int_0^{\infty}d\tau P^{-1/2-K}_{-1/2+i\tau}(\chi)P^{-1/2-K}_{-1/2+i\tau}(\chi)$ that involve these modes.  We shall present two approaches, one based on the integral relations given in (\ref{1.6}), and the other based on the complex $\tau$ plane structure.  In particular, we shall take advantage of the fact that, just as in (\ref{1.16}), it is possible to decompose non-singular functions such as  the sinc function $[\sin\tau]/\tau$ into a combination of two terms ($e^{i\tau}/2i\tau$ and $-e^{-i\tau}/2i\tau$) each one of which is singular at $\tau=0$.  Our key result will be to use (\ref{1.6}) in order to write the general  $P^{-1/2-K}_{-1/2+i\tau}(\chi)$ in terms of sinc functions according to
\begin{align}
P^{-1/2-K}_{-1/2+i \tau}(\chi)&=2\sum_{n=-\infty}^{\infty}\mathcal{R}^{K}_n(\chi)\frac{\sin[(\tau-in)\chi]}{(\tau-in)},
\label{1.18}
\end{align}
where
\begin{align}
\mathcal{R}^{K}_{n}(\chi)&=\frac{1}{(2\pi)^{3/2}\tanh^{K}\chi\sinh^{1/2}\chi\Gamma(1+K)}\int_{0}^{2\pi} d\omega \left (1-\frac{\cos \omega}{\cosh\chi}\right)^{K}e^{in\omega},
\label{1.19}
\end{align}
and where $n$ increases in unit steps, With  integer, non-negative $\ell$ (\ref{1.18}) takes the form
\begin{align}
P^{-1/2-\ell}_{-1/2+i \tau}(\chi)&=2\sum_{n=-\ell}^{\ell}\mathcal{R}^{\ell}_n(\chi)\frac{\sin[(\tau-in)\chi]}{(\tau-in)}=2\sum_{n=0}^{2\ell}\mathcal{R}^{\ell}_{n-\ell}(\chi)\frac{\sin[(\tau-i(n-\ell))\chi]}{(\tau-i(n-\ell))},
\label{1.20}
\end{align}
where
\begin{align}
\mathcal{R}^{\ell}_{n}(\chi)&=\frac{1}{(2\pi)^{3/2}\tanh^{\ell}\chi\sinh^{1/2}\chi\Gamma(1+\ell)}\int_{0}^{2\pi} d\omega \left (1-\frac{\cos \omega}{\cosh\chi}\right)^{\ell}e^{in\omega}.
\label{1.21}
\end{align}
That (\ref{1.18}) with its infinite sum on $n$ recovers (\ref{1.20}) with its finite sum on $n$ when $K=\ell$ is because, as we show below, for integer $n>\ell $ or $n<-\ell$ the function $\mathcal{R}^{\ell}_n(\chi)$ vanishes. On comparing (\ref{1.9})  with (\ref{1.20}), we see that  for practical numerical purposes it is easier to work with  a sum of $2\ell+1$ terms, rather than $\ell+1$ successive derivatives, especially in cosmology where the $\ell$ dependence of the anisotropy of the cosmic microwave background has been measured out to $\ell=1000$ or so.

Similarly for $Q^{-1/2-K}_{-1/2+i \tau}(\chi)$ with integer or non-integer $K$ use of (\ref{1.6}) leads us to
\begin{align} 
Q^{-1/2-\ell}_{-1/2+i   \tau}(\chi)&=-\pi\sum_{n=-\ell}^{\ell}\mathcal{R}^{\ell}_{n}(\chi)\frac{e^{-i   \chi(\tau-i    n)}}{\tau-i    n}=-\pi\sum_{n=0}^{2\ell}\mathcal{R}^{\ell}_{n-\ell}(\chi)\frac{e^{-i   \chi(\tau-i (n-\ell))}}{(\tau-i (n-\ell))},
\nonumber\\
Q^{-1/2-K}_{-1/2+i\tau}(\chi)&=-\pi\sum_{n=0}^{\infty}\mathcal{R}^K_{n-K}(\chi)
\frac{e^{-i \chi(\tau-i(n-K))}}{(\tau-i(n-K))},
\label{1.22}
\end{align}   
with the same $\mathcal{R}^{\ell}_{n}(\chi)$ when $K=\ell$ as in the $P^{-1/2-\ell}_{-1/2+i \tau}(\chi)$ case, as must be the case because of the  connection formula  that follows from (\ref{1.16}), viz. 
\begin{align}
\pi P^{-1/2-\ell}_{-1/2+i\tau}(\chi)=&-i[Q^{-1/2-\ell}_{-1/2+i\tau}(\chi)+Q^{-1/2-\ell}_{-1/2-i\tau}(\chi)],
\label{1.23}
\end{align}
a relation that directly exhibits the compatibility of (\ref{1.20}) and (\ref{1.22}) when $K=\ell$.

Since $\cos\omega$ is always less then  $\cosh\chi$ if $\chi>0$, the $\int_0^{2\pi}d\omega(1-\cos\omega/\cosh\chi)^K$ integral is never divergent no matter what the value of $K$. Thus even though we derived (\ref{1.18}), (\ref{1.20}) and (\ref{1.22}) starting from integral representations given in (\ref{1.6}) that were only finite under certain constraints, the relations given in (\ref{1.18}), (\ref{1.20}) and (\ref{1.22}) hold independent of those constraints. The ${\rm Re}[K]>1$ constraint when  $\omega=\chi$ is avoided since $\int_0^{2\pi}d\omega(1-\cos\omega/\cosh\chi)^K$ is finite even if ${\rm Re}[K]$ is less than or equal to $-1$. 

To understand what happens to the divergence at $\omega=\infty$ in the integral representation $\int_{\chi}^{\infty}d\omega  e^{-i\omega \tau}(\cosh\omega /\cosh K-1)^{K}$of $Q^{-1/2-K}_{-1/2+i\tau}(\chi)$ that is given in (\ref{1.6}), we note that the poles in $Q^{-1/2-K}_{-1/2+i\tau}(\chi)$ as given in (\ref{1.22}) are at $i\tau=K-n$ with $n \geq 0$. The large $\omega$ behavior of  $ e^{-i\omega \tau}(\cosh\omega /\cosh K-1)^{K}$  is of the form $e^{-i\omega \tau+\omega K}/\cosh^K\chi$, and at the poles $e^{-i\omega \tau+\omega K}=e^{\omega n}$, to thus not converge at these $n \geq 0$ poles. I.e.,  $Q^{-1/2-K}_{-1/2+i\tau}(\chi)$ diverges at its poles because $\int_{\chi}^{\infty}d\omega e^{-i\omega \tau}(\cosh\omega /\cosh K-1)^{K}$ diverges at them. (While $e^{\omega n}$ would not converge for any non-negative $n$ integer or not, the non-integer $n$ are not associated with solutions to (\ref{1.2}).)  Thus once we have extracted the pole terms in (\ref{1.22}), whatever is left over in $\mathcal{R}^K_{n-K}$ must be finite, just as $\int_0^{2\pi}d\omega e^{i(n-K)\omega}(1-\cos\omega/\cosh\chi)^K$  indeed is. Thus even though $\int_{\chi}^{\infty}d\omega  e^{-i\omega \tau}(\cosh\omega /\cosh \chi-1)^{K}$ diverges at the poles of $Q^{-1/2-K}_{-1/2+i\tau}(\chi)$, we can use the integral to  characterize the behavior at those poles.

As derived, all of the relations given in (\ref{1.18}), (\ref{1.20}) and (\ref{1.22}) isolate the $\tau$ dependence, to thereby enable us to obtain closed form expressions for integrals over $\tau$ of  associated Legendre conical functions and their products as discussed below, with a typical one being
\begin{align}
 \int_{-\infty}^{\infty}d \tau P^{-1/2-A}_{-1/2+i \tau}(\chi)P^{-1/2-B}_{-1/2-i \tau}(\chi)  =\left(\frac{2\pi}{\sinh\chi}\right)^{1/2}\frac{\Gamma (1+A+B)}{\Gamma (1+A)\Gamma (1+B)}P^{-1/2-(A+B)}_{-1/2}(\chi),
 \label{1.24}
\end{align}
for general $A$ and $B$.
As such, this paper is a follow up to a previous associated Legendre function  study of ours \cite{Liu2024}.

\section{Using the integral relations for the conical functions of the first kind}
\label{S2}

\subsection{$P^{-1/2-\ell}_{-1/2+i\tau}(\chi)$ functions with positive integer $\ell$}
\label{S2.1}

For $K$ equal to a positive integer $\ell$  we can expand the integrand in (\ref{1.6}) as 
\begin{align}
\left(1-\frac{\cosh \omega}{\cosh\chi}\right)^{\ell}=\sum_{m=0}^{\ell}(-1)^{m}\frac{\ell!}{m!(\ell-m)!}\frac{\cosh^{m}\omega}{\cosh^{m}\chi}= \sum_{m=0}^{\ell}(-1)^{m}\frac{\ell!}{m!(\ell-m)!}\frac{(e^{\omega}+e^{-\omega})^{m}}{2^m\cosh^{m}\chi}.
\label{2.1}
\end{align}
We can thus  write the integrand  in (\ref{1.6}) as an expansion in integer $n$ powers of $e^{\pm n\omega}$ for $-\ell\leq n\leq\ell$ with coefficients $\mathcal{R}_{n}^{\ell}(\chi)$:
\begin{align}
 \frac{1}{(2\pi)^{1/2}}\frac{\tanh^{-\ell}\chi}{\sinh^{1/2}\chi}\frac{1}{\Gamma(1+\ell)}\left (1-\frac{\cosh \omega}{\cosh\chi}\right)^{\ell}=\sum_{n=-\ell}^{\ell}\mathcal{R}^{\ell}_{n}(\chi)e^{-n\omega}.
 \label{2.2}
 \end{align}
Given (\ref{1.6}),  integrating (\ref{2.2}) over $\omega$ yields
\begin{align}
P^{-1/2-\ell}_{-1/2+i \tau}(\chi)=\sum_{n=-\ell}^{\ell}\mathcal{R}^{\ell}_{n}(\chi)\int_{-\chi}^{\chi}d\omega e^{-n\omega}e^{-i\omega\tau}=2\sum_{n=-\ell}^{\ell}\mathcal{R}^{\ell}_{n}(\chi)\frac{\sin((\tau-i n)\chi)}{(\tau-i n)}=2\sum_{n=0}^{2\ell}\mathcal{R}^{\ell}_{n-\ell}(\chi)\frac{\sin((\tau-i (n-\ell))\chi)}{(\tau-i (n-\ell))}.
\label{2.3}
\end{align}
On  setting  $\omega\rightarrow i \omega$,  (\ref{2.2}) takes the form 
\begin{align}
 \frac{1}{(2\pi)^{1/2}}\frac{\tanh^{-\ell}\chi}{\sinh^{1/2}\chi}\frac{1}{\Gamma(1+\ell)}\left (1-\frac{\cos \omega}{\cosh\chi}\right)^{\ell}=\sum_{n=-\ell}^{\ell}\mathcal{R}^{\ell}_{n}(\chi)e^{-in\omega}.
 \label{2.4}
 \end{align}
Using  the orthogonality of $e^{-i  n\omega}$ functions we obtain 
\begin{align}
\mathcal{R}^{\ell}_{n}(\chi)=\frac{1}{2\pi}\left(\frac{1}{2\pi}\right)^{1/2}\frac{\tanh^{-\ell}\chi}{\sinh^{1/2}\chi}\frac{1}{\Gamma(1+\ell)}\int_{0}^{2\pi} d\omega \left (1-\frac{\cos \omega}{\cosh\chi}\right)^{\ell}e^{in\omega}.
\label{2.5}
\end{align}
From (\ref{2.5}) it follows that $\mathcal{R}^{\ell}_{n}(\chi)=\mathcal{R}^{\ell}_{-n}(\chi)$. Thus finally we can write various forms for $P^{-1/2-\ell}_{-1/2+i \tau}(\chi)$, viz. 
\begin{align}
P^{-1/2-\ell}_{-1/2+i \tau}(\chi)&=\sum_{n=-\ell}^{\ell}\frac{1}{\pi}\left(\frac{1}{2\pi}\right)^{1/2}\frac{\tanh^{-\ell}\chi}{\sinh^{1/2}\chi}\frac{1}{\Gamma(1+\ell)}\frac{\sin((\tau-i n)\chi)}{(\tau-i n)}\int_{0}^{2\pi} d\omega \left (1-\frac{\cos \omega}{\cosh\chi}\right)^{\ell}e^{in\omega}
\nonumber\\
&=\frac{2R^{\ell}_0(\chi) \sin(\tau \chi)}{\tau}+\sum _{n=1}^{\ell}\frac{4R^{\ell}_n(\chi)}{(\tau^2+n^2)}[\tau\sin(\tau\chi)\cosh(n\chi)+n\cos(\tau\chi)\sinh(n \chi )].
\label{2.6}
\end{align}

Given (\ref{1.7})  and the (\ref{1.12}) recurrence relation we can immediately check the validity of (\ref{2.6})  for $\ell=0$, $\ell=1$ and $\ell=2$ \cite{footnote5}. In addition, we note that the integral $I(n,\ell)=\int_0^{2\pi}d \omega e^{in\omega}(1-\cos\omega/\cosh\chi)^{\ell}$ in (\ref{2.6}) with integer $n$ and $\ell$ is zero if $n>\ell$ or $n<-\ell$. Specifically,   with integer $n$ and $m$  we have
\begin{align}
&\int_0^{2\pi}d \omega e^{in\omega}\cos^m\omega=\int_0^{2\pi}d \omega e^{in\omega}\frac{e^{-im\omega}}{2^m}(1+e^{2i\omega})^m\sim\int_0^{2\pi}d \omega e^{i(n-m)\omega}
\left[1+e^{2i\omega}+e^{4i\omega}+.......e^{2im\omega}\right].
\label{2.7}
\end{align}
Since $\int_0^{2\pi}d \omega e^{ia\omega}$ with integer $a$ is only non-zero if $a=0$, it follows that $\int_0^{2\pi}d \omega e^{in\omega}\cos^m\omega$ is only non-zero if $n-m=(0,-2, ...,-2m)$, i.e., $n=(m,m-2, ...,-m)$, to thus vanish if $n>m$ or $n<-m$. Since the highest power of $\cos\omega$ in $I(n,\ell)$ is $\cos^{\ell}\omega$, it follows that  $I(n,\ell)$ vanishes if $n>\ell$ or $n<-\ell$.

In (\ref{2.6}) we have conveniently written $P^{-1/2-\ell}_{-1/2+i \tau}(\chi)$ in  a form in which the entire $\tau$ behavior is isolated in  sinc functions. This will prove to be very useful in the following.

\subsection{$P^{-1/2-K}_{-1/2+i \tau}(\chi)$ functions with non-integer ${\rm Re}[K]>0$}
\label{S2.2}

Writing $z=(\cosh\omega/\cosh\chi)$ where $|z|<1$ in the range $-\chi\leq  \omega\leq \chi$ associated with the integral range given in (\ref{1.6}), we can expand $(1-z)^{K}$ around $z=0$. For non-integer  $K$ the expansion gives an infinite binomial series that converges for $|z|<1$ viz. 
\begin{align}
& \left(1-\frac{\cosh \omega}{\cosh\chi}\right)^{K}=\sum_{n=0}^{\infty}\frac{(-1)^{n}}{n!}\frac{\Gamma (K+1)}{\Gamma (K+1-n)}\left(\frac{\cosh\omega}{\cosh\chi}\right)^{n},\qquad
{[}-\chi<\omega<\chi{]},
\label{2.8}
\end{align}
where $n$ is an integer that increases in unit steps.
We can then expand $\cosh^{n}\omega$ in powers of $e^{\pm n\omega}$ and then integrate over $\omega$ to get the series
\begin{align}
P^{-1/2-K}_{-1/2+i \tau}(\chi)=2\sum_{n=-\infty}^{\infty}\mathcal{R}^{K}_{n}(\chi)\frac{\sin(\chi(\tau-i  n))}{(\tau-i  n)}, 
\label{2.9}
\end{align}
a series that  for non-integer  $K$ has an  infinite number of terms.  Finally, with 
\begin{align}
& \frac{1}{(2\pi)^{1/2}}\frac{\tanh^{-K}\chi}{\sinh^{1/2}\chi}\frac{1}{\Gamma(1+K}\left (1-\frac{\cosh \omega}{\cosh\chi}\right)^{K}=\sum_{n=-\infty}^{\infty}\mathcal{R}^{K}_{n}(\chi)e^{-n\omega},
 \nonumber\\
 &\mathcal{R}^{K}_{n}(\chi)=\frac{1}{2\pi}\left(\frac{1}{2\pi}\right)^{1/2}\frac{\tanh^{-K}\chi}{\sinh^{1/2}\chi}\frac{1}{\Gamma(1+K)}\int_{0}^{2\pi}d \omega\left(1-\frac{\cos\omega}{\cosh\chi}\right)^{K}e^{i  \omega n},
\label{2.10}
\end{align}
we can write
\begin{align}
P^{-1/2-K}_{-1/2+i \tau}(\chi)=\sum_{n=-\infty}^{\infty}\frac{1}{\pi}\left(\frac{1}{2\pi}\right)^{1/2}\frac{\tanh^{-K}\chi}{\sinh^{1/2}\chi}\frac{1}{\Gamma(1+K)}\frac{\sin(\chi(\tau-i  n))}{(\tau-i  n)}\int_{0}^{2\pi} d\omega \left (1-\frac{\cos \omega}{\cosh\chi}\right)^{K}e^{i\omega n}.
\label{2.11}
\end{align}
With its infinite summation (\ref{2.11}) extends (\ref{2.6}) to non-integer $\ell$, with the $\tau$ dependence again being isolated. We note also that regardless of the value of  $K$, the $\omega$ integral in (\ref{2.11}) is always finite since the integration range is finite and $\cos \omega/\cosh\chi$ is always less than one if $\chi>0$.  Thus even though the $P^{-1/2-K}_{-1/2+i \tau}(\chi)$ integral given in (\ref{1.6}) required that ${\rm Re}[K]>-1$, (\ref{2.11}) poses no such constraint. Thus even though we used the $P^{-1/2-K}_{-1/2+i \tau}(\chi)$ integral in order to derive (\ref{2.11}), we can consider (\ref{2.11}) to be valid in its own right. (Essentially, once we derive (\ref{2.11}) starting with a specific set of constraints on $K$ we can then analytically continue to other values of $K$, since for any $K$ the $P^{-1/2-K}_{-1/2+i \tau}(\chi)$ functions have no complex $\tau$ plane singularities.)

\subsection{The $K-\ell$ Expansion for the $P^{-1/2-K}_{-1/2+i \tau}(\chi)$ functions}

Since the $P^{-1/2-\ell}_{-1/2+i \tau}(\chi)$ with non-negative, integer $\ell$ are complete, it must be  is possible to write the fluctuations with non-integer  $K$ as a sum over the basis of fluctuations with integer $\ell$. As before we define $z=\cosh\omega/\cosh\chi$ and expand the integrand $(1-z)^{K}$ in powers of $z^{n}$. We can then write $z^{n}=(1-(1-z))^{n}$ and expand in powers of $(1-z)^{\ell}$. With both expansions and with $-\chi \leq \omega \leq  \chi$ we obtain
\begin{align}
{\bigg (}1-\frac{\cosh \omega}{\cosh\chi}{\bigg )}^{K}&=\sum_{n=0}^{\infty}\frac{(-1)^{n}}{n!}\frac{\Gamma (1+K)}{\Gamma (1+K-n)}{\bigg (}\frac{\cosh\omega}{\cosh\chi}{\bigg )}^{n}
\nonumber\\ 
&= \sum_{n=0}^{\infty}\frac{(-1)^{n}}{n!}\frac{\Gamma (K+1)}{\Gamma (K+1-n)}{\bigg [}\sum_{\ell=0}^{n}\frac{(-1)^{\ell}}{\ell!}\frac{n!}{(n-\ell)!}{\bigg (}1-\frac{\cosh\omega}{\cosh\chi}{\bigg )}^{\ell}{\bigg ]}.
\label{2.12}
\end{align}
This allows us to write the $P^{-1/2-K}_{-1/2+i \tau}(\chi)$ with non-integer  $K$ in terms of a double sum over the $P^{-1/2-\ell}_{-1/2+i \tau}(\chi)$ functions with integer $\ell$, viz.
\begin{align}
P^{-1/2-K}_{-1/2+i \tau}(\chi)=\sum_{n=0}^{\infty}\sum_{\ell=0}^{n}\frac{(-1)^{n+\ell}}{(n-\ell)!}\frac{\tanh^{\ell-K}\chi}{\Gamma (1+K-n)} P^{-1/2-\ell}_{-1/2+i \tau}(\chi).
\label{2.13}
\end{align}

\subsection{Products of $P^{-1/2-K}_{-1/2+i \tau}(\chi)$ functions}

As shown in (\ref{1.13}), for integer $\ell$ we can  form a $\chi$-space orthonormal basis in which we integrate over $\chi$. We now show that we can determine integrals over $\tau$ of products of $P^{-1/2-K}_{-1/2+i \tau}(\chi)$  and $P^{-1/2-K}_{-1/2-i \tau}(\chi)$ functions  in a closed form. From this we could then construct an orthonormal (in $\tau$) basis using a Gram–Schmidt process. To be specific, using the integral representations given in (\ref{1.6}) of two $P^{-1/2-K}_{-1/2+i \tau}(\chi)$ functions  with general parameters $A$ and $B$ that satisfy ${\rm Re}[A]>-1$, ${\rm Re}[B]>-1$, with $P^{-1/2-K}_{-1/2+i \tau}(\chi)=P^{-1/2-K}_{-1/2-i \tau}(\chi)$ we obtain
\begin{align}
&P^{-1/2-A}_{-1/2+i \tau}(\chi)P^{-1/2-B}_{-1/2-i \tau}(\chi)
\nonumber\\
& =\frac{1}{2\pi}\frac{\tanh^{-A-B}\chi}{\sinh\chi}\frac{1}{\Gamma (1+A)\Gamma (1+B)}\int_{-\chi}^{\chi}d \omega\int_{-\chi}^{\chi} d \omega^{\prime}{\bigg (}1-\frac{\cosh\omega}{\cosh\chi}{\bigg )}^{A}{\bigg (}1-\frac{\cosh\omega^{\prime} }{\cosh\chi}{\bigg )}^{B}e^{-i (\omega-\omega^{\prime} )\tau}, 
\label{2.14}
\end{align}
Integrating over $\tau$ gives  $2\pi\delta(\omega-\omega^{\prime} )$, so that integrating over $\omega^{\prime} $ then yields
\begin{align}
 \int_{-\infty}^{\infty} d\tau P^{-1/2-A}_{-1/2+i \tau}(\chi)P^{-1/2-B}_{-1/2-i \tau}(\chi)  = \frac{\tanh^{(-A-B)}\chi}{\sinh\chi}\frac{1}{\Gamma (1+A)\Gamma (1+B)}\int_{-\chi}^{\chi}d \omega  {\bigg (}1-\frac{\cosh\omega}{\cosh\chi}{\bigg )}^{A+B}.
 \label{2.15}
 \end{align}
We recognize the right-hand side of (\ref{2.15}) as containing none other than a $P^{-1/2-K}_{-1/2+i \tau}(\chi)$ function with $\tau=0$, so that
\begin{align}
 \int_{-\infty}^{\infty}d \tau P^{-1/2-A}_{-1/2+i \tau}(\chi)P^{-1/2-B}_{-1/2-i \tau}(\chi)  =\left(\frac{2\pi}{\sinh\chi}\right)^{1/2}\frac{\Gamma (1+A+B)}{\Gamma (1+A)\Gamma (1+B)}P^{-1/2-(A+B)}_{-1/2}(\chi).
 \label{2.16}
\end{align}
As a check on (\ref{2.16}) we set  $A=0$, $B=0$ and obtain 
\begin{align}
& \int_{-\infty}^{\infty}d \tau P^{-1/2}_{-1/2+i \tau}(\chi)P^{-1/2}_{-1/2-i \tau}(\chi)  =\frac{2}{\pi \sinh\chi}\int_{-\infty}^{\infty} d\tau \frac{\sin^2(\tau\chi)}{\tau^2}=\frac{2\chi}{\pi \sinh\chi}\int_{-\infty}^{\infty} d\tau \frac{\sin^2\tau}{\tau^2}=\frac{2\chi}{ \sinh\chi},
 \nonumber\\
 &\left(\frac{2\pi}{\sinh\chi}\right)^{1/2}\frac{\Gamma (1)}{\Gamma (1)\Gamma (1)}P^{-1/2}_{-1/2}(\chi)=\frac{2\chi}{ \sinh\chi}.
 \label{2.17}
\end{align}
We thus confirm (\ref{2.16}) when $A=0$, $B=0$. 

We note that while we derived (\ref{2.16})  in the domain where both of the $P^{-1/2-A}_{-1/2+i \tau}(\chi)$ and $P^{-1/2-B}_{-1/2-i \tau}(\chi)$ functions  that appear on its left-hand side  have an integral representation, viz. ${\rm Re}[A]>-1$ and ${\rm Re}[B]>-1$,  and in which the integral on the right-hand side of (\ref{2.15})  requires ${\rm Re}[A+B]>-1$, the equality analytically extends beyond this domain to everywhere where the integral over $\tau$ of the $P^{-1/2-A}_{-1/2+i \tau}(\chi)P^{-1/2-B}_{-1/2-i \tau}(\chi)$ product  converges. In Sec. \ref{S4} we will show how to extend this analysis to $\tau$ integrals of more than two $P^{-1/2-K}_{-1/2+i \tau}(\chi)$.

We also note that on introducing  constants $a$ and $b$, via use of (\ref{1.6}) (\ref{2.16}) has a straightforward generalization:
\begin{align}
&\int_{-\infty}^{\infty}d \tau P^{-1/2-A}_{-1/2+i (\tau+a)}(\chi)P^{-1/2-B}_{-1/2-i (\tau+b)}(\chi)
\nonumber\\
&=\frac{\tanh^{-(A+B)}\chi}{2\pi \sinh\chi\Gamma(1+A)\Gamma(1+B)}\int_{-\infty}^{\infty}d \tau\int_{-\chi}^{\chi}d\omega \int_{-\chi}^{\chi} d\omega^{\prime}{\bigg (}1-\frac{\cosh\omega}{\cosh\chi}{\bigg )}^{A}{\bigg (}1-\frac{\cosh\omega^{\prime}}{\cosh\chi}{\bigg )}^{B}e^{-i(\omega-\omega^{\prime})\tau}e^{-ia\omega}e^{i b\omega^{\prime}}
\nonumber\\
&=\left(\frac{2\pi}{\sinh\chi}\right)^{1/2}\frac{\Gamma (1+A+B)}{\Gamma (1+A)\Gamma (1+B)}P^{-1/2-(A+B)}_{-1/2+i(a-b)}(\chi),
 \label{2.18}
\end{align}
in which we have made separate shifts in $\tau$ in each of the two conical functions.

\section{Using the integral relations for the conical functions of the second kind}
\label{S3}

\subsection{$Q^{-1/2-\ell}_{-1/2+i\tau}(\chi)$ functions   with positive integer $\ell$}
\label{S3.1}

With the integral relations for the $Q^{-1/2-\ell}_{-1/2+i\tau}(\chi)$ functions in (\ref{1.6}) containing the same $(1-\cosh\omega/\cosh \chi)^{\ell}$ factor as the $P^{-1/2-\ell}_{-1/2+i\tau}(\chi)$, (\ref{2.1}) again holds. However, now we have to integrate over an infinite range for $\omega$. As with (\ref{1.6}) and (\ref{1.7}) where we had to consider $Q^{-1/2}_{-1/2+i(\tau-i\epsilon)}(\chi)$ in order to get convergence at large $\omega$, for general $Q^{-1/2-K}_{-1/2+i(\tau-i\epsilon)}(\chi)$ we require ${\rm Im}[\tau]<-{\rm Re}[K]$, and for such $\tau$ obtain 
\begin{align} 
 i   \int_{\chi}^{\infty}d\omega e^{-i   \omega(\tau-i    n)}=\frac{e^{-i   \chi(\tau-i    n)}}{\tau-i    n}, 
 \label{3.1}
\end{align}  
provided $n-\tau_I>0$. 
Other than this one difference in which (\ref{3.1}) replaces the integral that appears  in (\ref{2.3}), everything goes through as with the $P^{-1/2-\ell}_{-1/2+i\tau}(\chi)$, to thus yield
\begin{align} 
Q^{-1/2-\ell}_{-1/2+i   \tau}(\chi)=-\pi\sum_{n=-\ell}^{\ell}\mathcal{R}^{\ell}_{n}(\chi)\frac{e^{-i   \chi(\tau-i    n)}}{\tau-i    n}=-\pi\sum_{n=0}^{2\ell}\mathcal{R}^{\ell}_{n-\ell}(\chi)\frac{e^{-i   \chi(\tau-i (n-\ell))}}{\tau-i (n-\ell)},
\label{3.2}
\end{align}   
where the $\mathcal{R}^{\ell}_{n}(\chi)$ are same functions as given in (\ref{2.5}) for $P^{-1/2-\ell}_{-1/2+i   \tau}(\chi)$. Given (\ref{1.7})  and (\ref{1.12}) we can immediately check the validity of this relation for $\ell=0$, $\ell=1$ and $\ell=2$. 
As with  (\ref{2.6}), in (\ref{3.2})  we have conveniently written $Q^{-1/2-\ell}_{-1/2+i \tau}(\chi)$ in  a form in which the entire $\tau$ behavior is isolated. Again we note also that regardless of the value of  $\ell$, the $\omega$ integral in $R^{\ell}_n$  is always finite since $\cos \omega/\cosh\chi$ is always less than one and the integration range is finite. Thus even though the $Q^{-1/2-\ell}_{-1/2+i \tau}(\chi)$ integral given in (\ref{1.6}) required that $-1<\ell<-{\rm Im}[\tau]$,  and even though the integration in (\ref{3.1}) required that $n> {\rm Im}[\tau]$ for  all $n$, so again $-\ell > {\rm Im}[\tau]$,  (\ref{3.2}) poses no such constraint. I.e., while the $Q^{-1/2-\ell}_{-1/2+i\tau}(\chi)$ integral given in (\ref{1.6}) diverges for certain complex values of $\tau$, by isolating these terms in (\ref{3.2}), the integral that appears in $\mathcal{R}^{\ell}_n(\chi)$ has to be, and as noted above, is in fact finite.

Thus even though we used the $Q^{-1/2-\ell}_{-1/2+i \tau}(\chi)$ integral in order to derive (\ref{3.2})  we can consider (\ref{3.2}) and (\ref{3.5}) below to be valid in their own right, and will thus derive (\ref{3.5}) below without constraint on $K$. Once we derive (\ref{3.2}) starting with a specific set of integer values for $K$ we can then analytically continue to other values of $K$, only now and unlike in  the $P^{-1/2-K}_{-1/2+i \tau}(\chi)$ case, for $Q^{-1/2-K}_{-1/2+i \tau}(\chi)$ with non-integer $K$ we will in (\ref{3.5}) encounter a new class of complex $\tau$ plane singularities of the type identified in \cite{Liu2024}. Specifically, it was noted there that while the number of complex $\tau$ plane poles is finite when $K$ is integer, when $K$ is not an integer the number is infinite. We now show this using the techniques developed in this paper. 

\subsection{$Q^{-1/2-K}_{-1/2+i\tau}(\chi)$ functions   with non-integer ${\rm Re}[K]>0$}
\label{S3.2}

With there being an infinite number of $\tau$ plane poles in  $Q^{-1/2-K}_{-1/2+i\tau}(\chi)$ when $K$ is not  an integer, we will not be able to expand $Q^{-1/2-K}_{-1/2+i\tau}(\chi)$ the same way that we expanded $Q^{-1/2-\ell}_{-1/2+i\tau}(\chi)$, as that would lead to a requirement that $n$ go all the way down to $-\infty$, with there then being no solution to $n> {\rm Im}[\tau]$ for  all $n$. Thus we must expand $Q^{-1/2-K}_{-1/2+i\tau}(\chi)$ so that the minimum value needed for $n$ is bounded from below, and shall seek to have the minimum needed value for $n$ be zero. To this end we note that for non-integer $K$ we had expanded the integrand in the $P^{-1/2-K}_{-1/2+i\tau}(\chi)$ integral around $\cosh\omega/\cosh \chi =0$. With the range of integration for the $Q^{-1/2-K}_{-1/2+i\tau}(\chi)$ integral in (\ref{1.6})  being such that $\cosh\omega/\cosh \chi >1$,  we instead first  pull out a factor of $e^{K\omega}$ to give 
\begin{align} 
{\bigg (}\frac{\cosh\omega}{\cosh\chi}-1{\bigg )}^{K}=\frac{e^{K\omega}}{2^{K}\cosh^{K}\chi}{\bigg (}1-e^{-\omega}{\big (}2\cosh\chi-e^{-\omega}{\big )}{\bigg )}^{K}.
\label{3.3}
\end{align}  
We introduce $\tilde{z}=e^{-\omega}{\big (}2\cosh\chi-e^{-\omega}{\big )}$, with  $\tilde{z}=0$ corresponding to $\omega=\infty$ and $\tilde{z}=1$ corresponding to $\omega=\chi$. With $\chi<\omega<\infty$ we have $|\tilde{z}|<1$. Since $\tilde{z}$ only contains factors of $e^{-\omega}$ and $e^{-2\omega}$ but no $e^{+\omega}$, the expansion is in terms of $e^{-n\omega}$ ranges from $0\leq n<\infty$, and is of the form
\begin{align} 
&-ie^{-i    K\pi}\left(\frac{\pi}{2}\right)^{1/2}\frac{\sinh^{-K}\chi}{\sinh^{1/2}\chi}\frac{e^{K\omega}}{2^{K}\Gamma(1+K)}{\bigg (}1-e^{-\omega}{\big (}2\cosh\chi-e^{-\omega}{\big )}{\bigg )}^{K}=\sum_{n=0}^{\infty}\tilde{\mathcal R}_{n}^{K}(\chi)e^{(K-n)\omega},
\label{3.4}
\end{align}
with the extracted factor  $e^{  K\omega}$ having led to  poles at $\tau=i   (K-n)$. Using (\ref{1.6}) we then obtain 
\begin{align}
&Q^{-1/2-K}_{-1/2+i(\tau-i\epsilon)}(\chi)=\int_{\chi}^{\infty}d\omega e^{-i\omega \tau}\sum_{n=0}^{\infty}\tilde{\mathcal R}_{n}^{K}(\chi)e^{(K-n)\omega}=-i\sum_{n=0}^{\infty}\tilde{\mathcal{R}}^{K}_{n}(\chi)\frac{e^{-i   \chi(\tau-i   (n-K))}}{\tau-i   (n-K)},\ \ \ \ {[}n-\tau_I+\epsilon>0{]},
\label{3.5}
\end{align} 
provided $n-\tau_I+\epsilon>0$.  With (\ref{3.5}) we have isolated the $\tau$ dependence of $Q^{-1/2-K}_{-1/2+i(\tau-i\epsilon)}(\chi)$.

In (\ref{3.2}) we have given two forms for  $Q^{-1/2-\ell}_{-1/2+i   \tau}(\chi)$ with integer $\ell$, the first one where $n$ begins at $-\ell$, and the second where we have shifted $n$ so that it begins at zero. Comparing the second form in (\ref{3.2}) with (\ref{3.5}) we see that $i\tilde{\mathcal{R}}^{\ell}_{n}(\chi)=\pi \mathcal{R}^{\ell}_{n-\ell}(\chi)$. For $K=0$ there is no need to shift $n$ in (\ref{3.2}), but for  for $K=1$  the first form given in   (\ref{3.2}) has poles at $n=-1$, $n= 0$, and $n=1$, whereas the corresponding poles in  the second form given in   (\ref{3.2}) are at $n=0$, $n= 1$, and $n=2$, just as they are in (\ref{3.5}) with the same $K=1$. A similar pattern occurs for all higher integer $K$, and is due to the fact that in deriving (\ref{3.5}) we had pulled out a factor of $e^{K\omega}$ in (\ref{3.3}), and this causes the range of poles in (\ref{3.5}) to begin at $n=0$.

The complex $\tau$ plane pole pattern that we have obtained can be read off from (\ref{1.17}). Specifically,  with $P^{-1/2-K}_{-1/2+i\tau}(\chi)$ and $P^{1/2+K}_{-1/2+i\tau}(\chi)$ having no complex $\tau$ plane poles, complex $\tau$ plane poles in $Q^{-1/2-K}_{-1/2+i\tau}(\chi)$ arise from the $\Gamma(i\tau-K)$ factor in (\ref{1.17}) unless cancelled by zeroes in the $1/\Gamma(i\tau+1+K)$ factor \cite{footnote6}. For integer $\ell$ both  poles and zeroes are relevant leading to the finite range $-\ell \leq n \leq \ell$ of poles given in (\ref{3.2}). However, for non-integer $K$ there is no cancellation of poles, and we get the semi-infinite set of poles with $0 \leq n \leq \infty$ that is exhibited in (\ref{3.5}). 

The complex $\tau$ plane pole pattern that we have obtained  for the $Q^{-1/2-K}_{-1/2+i\tau}(\chi)$ can also be exhibited \cite{Liu2024} by looking at the large $\chi$ limit of (\ref{1.3}), viz. 
\begin{align}
Q^{-1/2-K}_{-1/2+i\tau}(\chi)\rightarrow -\frac{i\pi^{1/2}e^{-i\pi K}\Gamma(i\tau-K)}{(2\cosh\chi)^{1/2+i\tau}\Gamma(i\tau+1)}\left(1+\frac{1+2K}{4\cosh^2\chi}\right)\left(1+\frac{(i\tau-K)(i\tau-K+1)}{4\cosh^2\chi(i\tau+1)}\right)+.....
\label{3.6}
\end{align}
The $\Gamma(i\tau-K)$  function has complex $\tau$ plane poles at $\tau-i(K-n)=0$, where $n=0,1,2,....$. This is just as exhibited in (\ref{3.5}), and entails an infinite number of poles if $K$ is not an integer, since in that case  the integer $n $ can take an infinite number of values in the $(0,\infty)$ range. However, the $1/\Gamma(i\tau+1)$ function has zeroes at $i\tau=-1,-2,-3,....$, and thus some or all of the poles in $\Gamma(i\tau-K)$ can be cancelled if $K$ is integer. As noted in \cite{Liu2024} this results in a quite unusual pattern. If $K$ is a negative integer there are no residual poles. When $K=0$ there is one residual pole, when $K=1$ there are three residual poles, when $K=2$ there are five residual poles, and so on \cite{footnote7}, even though there is an infinite number of poles if $K$ is not an integer, no matter how close to an integer $K$ may be. That there is only a finite number of poles when $K$ is an integer can also be seen from (\ref{3.2}), since as we had noted above $\mathcal{R}^{\ell}_n(\chi)$ is zero if $n>\ell$, $n<-\ell$.

To get an explicit expression for $Q^{-1/2-K}_{-1/2+i(\tau-i\epsilon)}(\chi)$ that exhibits this pole structure,   we multiply both sides of (\ref{3.4}) by  $e^{-K\omega}$, set $\omega \rightarrow i\omega$, and obtain
\begin{align}
&-i\left(\frac{\pi}{2}\right)^{1/2}\frac{\tanh^{-K}\chi}{\sinh^{1/2}\chi}\frac{1}{\Gamma(1+K)}\bigg{(}1-\frac{\cosh\omega}{\cosh\chi}\bigg {)}^{K}e^{-K\omega}=\sum_{n=0}^{\infty}\tilde{\mathcal R}_{n}^{K}(\chi)e^{-n\omega},
\nonumber\\
&-i\left(\frac{\pi}{2}\right)^{1/2}\frac{\tanh^{-K}\chi}{\sinh^{1/2}\chi}\frac{1}{\Gamma(1+K)}\bigg{(}1-\frac{\cos\omega}{\cosh\chi}\bigg {)}^{K}e^{-iK\omega}=\sum_{n=0}^{\infty}\tilde{\mathcal R}_{n}^{K}(\chi)e^{-in\omega},
\nonumber\\
&-\frac{i}{2\pi}\left(\frac{\pi}{2}\right)^{1/2}\frac{\tanh^{-K}\chi}{\sinh^{1/2}\chi}\frac{1}{\Gamma(1+K)}\int_0^{2\pi} d\omega\bigg{(}1-\frac{\cos\omega}{\cosh\chi}\bigg {)}^{K}e^{i(n-K)\omega}=\tilde{\mathcal R}_{n}^{K}(\chi).
\label{3.7}
\end{align}
Thus finally we obtain
\begin{align}
&Q^{-1/2-K}_{-1/2+i(\tau-i\epsilon)}(\chi)=-\sum_{n=0}^{\infty}\left[\frac{1}{2\pi}\left(\frac{\pi}{2}\right)^{1/2}\frac{\tanh^{-K}\chi}{\sinh^{1/2}\chi}\frac{1}{\Gamma(1+K)}\int_0^{2\pi} d\omega\bigg{(}1-\frac{\cos\omega}{\cosh\chi}\bigg {)}^{K}e^{i(n-K)\omega}\right]
\frac{e^{-i \chi(\tau-i(n-K))}}{(\tau-i(n-K))}.
\label{3.8}
\end{align}
Comparing with (\ref{2.10}) we again obtain $i\tilde{\mathcal{R}}^K_n=\pi\mathcal{R}^K_{n-K}$. Despite its derivation with constraints on ${\rm Im}[\tau]$ and ${\rm Re}[K]$  the expression for $Q^{-1/2-K}_{-1/2+i(\tau-i\epsilon)}(\chi)$ given in (\ref{3.8}) holds for all $\tau$ and all $K$, since the  $\tau$ pole dependence has been isolated and the $\omega$ integral is  finite for all $K$ since $1-\cos\omega/\cosh \chi$ with $\chi>0$ can never be zero. In order to recover (\ref{3.2}) from (\ref{3.8}) when $K$ is an integer $\ell$ we set $m=n-\ell$, with $m$ then ranging from $-\ell$ to $\infty$. However, with $i\tilde{\mathcal{R}}^{\ell}_n=\pi\mathcal{R}^{\ell}_{n-\ell}=\pi\mathcal{R}^{\ell}_{m}$, then  since $\mathcal{R}^{\ell}_{m}$ is only non-zero if $m$ is constrained to the range $-\ell \leq m\leq \ell$
(\ref{3.2}) follows just as it should. With (\ref{3.8}) we have obtained the  closed form expression for $Q^{-1/2-K}_{-1/2+i(\tau-i\epsilon)}(\chi)$ that we seek.

\subsection{The $K-\ell$ expansion for $Q^{-1/2-K}_{-1/2+i\tau}(\chi)$ functions  }
\label{S3.3}

As with the $P^{-1/2-K}_{-1/2+i \tau}(\chi)$, we can express the $Q^{-1/2-K}_{-1/2+i \tau}(\chi)$ with non-integer  $K$ in terms of $Q^{-1/2-\ell}_{-1/2+i \tau}(\chi)$ functions  with integer $\ell$. To this end we start with 
\begin{align}
 Q^{-1/2-K}_{-1/2+i\tau}(\chi)=-\frac{ie^{-i K\pi}}{2^{K}}\left(\frac{\pi}{2\sinh\chi}\right)^{1/2}\frac{\sinh^{-K}\chi}{\Gamma(1+K)}\int_{\chi}^{\infty} {\bigg (}1-e^{-\omega}{\big (}2\cosh\chi-e^{-\omega}{\big )}{\bigg )}^{K}e^{-i(\tau+i K)\omega}d\omega,
 \label{3.9}
 \end{align}
 and make the shift  $\tau\rightarrow \tau-iK$, to obtain 
\begin{align}
 Q^{-1/2-K}_{-1/2+K+i\tau}(\chi)=-\frac{ie^{-i K\pi}}{2^{K}}\left(\frac{\pi}{2\sinh\chi}\right)^{1/2}\frac{\sinh^{-K}\chi}{\Gamma(1+K)}\int_{\chi}^{\infty} {\bigg (}1-e^{-\omega}{\big (}2\cosh\chi-e^{-\omega}{\big )}{\bigg )}^{K}e^{-i\tau\omega}d\omega.
 \label{3.10}
 \end{align}
Next, we expand 
\begin{align} 
&{\bigg (}1-e^{-\omega}{\big (}2\cosh\chi-e^{-\omega}{\big )}{\bigg )}^{K}=\sum_{n=0}^{\infty}\frac{(-1)^{n}}{n!}\frac{\Gamma(1+K)}{\Gamma(1+K-n)}{\bigg (}e^{-\omega}{\big (}2\cosh\chi-e^{-\omega}{\big )}{\bigg )}^{n}
\nonumber\\ 
&= \sum_{n=0}^{\infty}\frac{(-1)^{n}}{n!}\frac{\Gamma(K+1)}{\Gamma(K+1-n)}{\bigg [}\sum_{\ell=0}^{n}\frac{(-1)^{\ell}}{\ell!}\frac{n!}{(n-\ell)!}{\bigg (}1-e^{-\omega}{\big (}2\cosh\chi-e^{-\omega}{\big )}{\bigg )}^{\ell}{\bigg ]}
\nonumber\\
&= \sum_{n=0}^{\infty}\sum_{\ell=0}^{n}\frac{\Gamma(K+1)}{\Gamma(K+1-n)}\frac{(-1)^{n+\ell}}{\ell!(n-\ell)!}{\bigg (}1-e^{-\omega}{\big (}2\cosh\chi-e^{-\omega}{\big )}{\bigg )}^{\ell}.
\label{3.11}
\end{align}
Finally,  recognizing that 
\begin{align}
 Q^{-1/2-\ell}_{-1/2+\ell+i\tau}(\chi)=-\frac{ie^{-i\ell\pi}}{2^{\ell}}\left(\frac{\pi}{2\sinh\chi}\right)^{1/2}\frac{\sinh^{-\ell}\chi}{\Gamma(1+\ell)}\int_{\chi}^{\infty} {\bigg (}1-e^{-\omega}{\big (}2\cosh\chi-e^{-\omega}{\big )}{\bigg )}^{\ell}e^{-i\tau\omega}d\omega,
 \label{3.12}
 \end{align}
we obtain the required
\begin{align}
Q^{-1/2-K}_{-1/2+K+i\tau}(\chi)= \frac{e^{-i K\pi}}{2^{K}\sinh^{K}\chi}\sum_{n=0}^{\infty}\sum_{\ell=0}^{n}\frac{(-1)^{n}}{\Gamma(K+1-n)}\frac{2^{\ell}\sinh^{\ell}\chi}{(n-\ell)!}Q^{-1/2-\ell}_{-1/2+\ell+i\tau}(\chi).
\label{3.13}
\end{align}

The shift $\tau\rightarrow \tau-iK$, $\tau\rightarrow \tau-i \ell$ was needed in order  to line up the relevant poles of $Q^{-1/2-K}_{-1/2+K+i\tau}(\chi)$ and $Q^{-1/2-\ell}_{-1/2+\ell+i\tau}(\chi)$ so that they are all  located at a common $\tau=i n$ with $n\geq 0$. For the $P^{-1/2-\ell}_{-1/2-i\tau}(\chi)$ and $P^{-1/2-K}_{-1/2-i\tau}(\chi)$ functions this was not needed since the relevant sinc functions are already lined up.

\subsection{Products of $Q^{-1/2-K}_{-1/2+i \tau}(\chi)$ functions}

In addition, we can construct an expression for the integral with respect to $\tau$ of the  $Q^{-1/2-A}_{-1/2+i\tau}(\chi)Q^{-1/2-B}_{-1/2-i\tau}(\chi)$ type products that is similar to the expression that we  found for the integral of the  $P^{-1/2-A}_{-1/2+i\tau}(\chi)P^{-1/2-B}_{-1/2-i\tau}(\chi)$ type products in (\ref{2.18}). While we could use (\ref{3.8}) and evaluate
$\int d\tau Q^{-1/2-A}_{-1/2+i\tau}(\chi)Q^{-1/2-B}_{-1/2-i\tau}(\chi)$ as a contour integral  we have found it more convenient to use the integral relation given in (\ref{1.6}).   Taking $K$ values $A$ and $B$ and introducing $a$ and $b$ shifts in  $\tau$ we obtain 
\begin{align}
&\int_{-\infty}^{\infty}   d\tau_R Q^{-1/2-A}_{-1/2+i(\tau_R+a-i\epsilon)}(\chi)Q^{-1/2-B}_{-1/2-i(\tau_R+b+i\epsilon)}(\chi)
\nonumber\\
&=-\frac{\pi e^{-i(A+B)\pi}\tanh^{-A-B}\chi}{2\sinh\chi\Gamma(1+A)\Gamma(1+B)}
 \int_{-\infty}^{\infty}  d\tau_R \int_{\chi}^{\infty}d\omega \int_{\chi}^{\infty} d\omega^{\prime}{\bigg (}\frac{\cosh\omega}{\cosh\chi}-1{\bigg )}^{A}{\bigg (}\frac{\cosh\omega^{\prime}}{\cosh\chi}-1{\bigg )}^{B}e^{-i(\omega-\omega^{\prime})\tau_R}e^{-ia\omega}e^{i b\omega^{\prime}},
\label{3.14}
\end{align}
where $\tau$ itself  is taken to be real  and set equal to $\tau_R$. 
We have introduced $a$ and $b$ here in order to secure  convergence on the real $\tau_R$ axis of the integral on the right-hand side of (\ref{3.14}). We recall that we were able to  use the integral relations in (\ref{1.6}) when $A=0$, $B=0$ if we used  $Q^{-1/2}_{-1/2+ i(\tau-i\epsilon)}(\chi)$ and $Q^{-1/2}_{-1/2- i(\tau+i\epsilon)}(\chi)$, and have thus introduced $i\epsilon$ factors here. Thus for convergence we require that $-{\rm Im}[a-i\epsilon]=\epsilon-{\rm Im}[a]>{\rm Re}[A]>-1$ and ${\rm Im}[b+i\epsilon]=\epsilon +{\rm Im}[b]>{\rm Re}[B]>-1$.

In (\ref{3.14}) we integrate  over $\tau_R$  and generate a $2\pi \delta(\omega-\omega^{\prime})$ term, and then integrate over $\omega^{\prime}$, and then again using (\ref{1.6})   obtain
\begin{align}
 \int_{-\infty}^{\infty} d\tau_R Q^{-1/2-A}_{-1/2+i(\tau_R+a-i\epsilon)}(\chi)Q^{-1/2-B}_{-1/2-i(\tau_R+b+i\epsilon)}(\chi) 
&=-\frac{\pi^2e^{-i(A+B)\pi}\tanh^{-A-B}\chi}{\sinh\chi\Gamma(1+A)\Gamma(1+B)}
\int_{\chi}^{\infty} d\omega{\bigg (}\frac{\cosh\omega}{\cosh\chi}-1{\bigg )}^{A+B}e^{-i (a-b-2i\epsilon)\omega}
\nonumber \\
&=-i \pi \left(\frac{2\pi}{\sinh\chi}\right)^{1/2}\frac{\Gamma(1+A+B)}{\Gamma(1+A)\Gamma(1+B)}Q^{-1/2-(A+B)}_{-1/2+i (a-b)+2\epsilon}(\chi).
\label{3.15}
\end{align}
Other than the presence of the $i\epsilon$ factors  (\ref{3.15}) is similar to (\ref{2.18}).

To illustrate the validity of  (\ref{3.15}) we evaluate it for  $A=0$, $B=0$.  However, while we should keep $a-b$ non-zero in order to get a finite answer since $Q^{-1/2}_{-1/2+ i\tau}(\chi) \sim 1/ \tau$, we can satisfy the $\epsilon-{\rm Im}[a]>0>-1$, $\epsilon +{\rm Im}[b]>0>-1$ conditions if we  set both $a$ and $b$ to zero and keep non-zero $\epsilon$. Then we obtain 
\begin{align}  
&\int_{-\infty}^{\infty} d\tau_R Q^{-1/2}_{-1/2+ i(\tau_R-i\epsilon)}(\chi)Q^{-1/2}_{-1/2- i(\tau_R+i\epsilon)}(\chi)
=-\frac{\pi}{2\sinh\chi}\int_{-\infty}^{\infty} d\tau_R \frac{e^{-2\chi \epsilon}}{(\tau_R-i\epsilon)(\tau_R+i\epsilon)}=-\frac{\pi}{2\sinh\chi}\frac{\pi e^{-2\chi \epsilon}}{\epsilon},
\label{3.16}
\end{align}  
\begin{align}  
&-i\pi\left(\frac{2\pi}{\sinh\chi}\right)^{1/2}Q^{-1/2}_{-1/2+i(-2i\epsilon)}(\chi)=-i\pi\frac{\pi}{\sinh\chi}\frac{e^{-2\chi\epsilon}}{2i\epsilon}=-\frac{\pi}{\sinh\chi}\frac{\pi e^{-2\chi \epsilon}}{2\epsilon}.
\label{3.17}
\end{align}  
Thus while both sides of (\ref{3.15}) diverge in this particular case, they diverge at exactly the same rate, and the integral relation in (\ref{3.15}) accommodates it.  Thus we can use the integral expressions in (\ref{1.6}) even if they do not lead us to expressions that converge. Now we had used (\ref{1.6}) to extract the $\tau$ plane pole structure of the $Q^{-1/2-K}_{-1/2+i \tau}(\chi)$ in the region where the (\ref{1.6}) integral did not converge, and in (\ref{3.8}) we had expanded the general $Q^{-1/2-K}_{-1/2+i \tau}(\chi)$ entirely in terms of $\tau$ plane poles, with (\ref{3.8}) holding regardless of whether or not the (\ref{1.6}) integral diverges.  With convergence thus not being an issue here  we can set $a=0$, $b=0$ in (\ref{3.15}), and obtain
\begin{align}  
&\int_{-\infty}^{\infty} d\tau_R Q^{-1/2-A}_{-1/2+i ( \tau_R-i\epsilon)}(\chi)Q^{-1/2-B}_{-1/2-i  (\tau_R+i\epsilon)}(\chi)
= -i    \pi \left(\frac{2\pi}{\sinh\chi}\right)^{1/2}\frac{\Gamma   (1+A+B)}{\Gamma   (1+A)\Gamma   (1+B)}Q^{-1/2-A-B}_{-1/2+2\epsilon}(\chi).
\label{3.18}
\end{align}  

As a check on (\ref{3.18}) we take $A=1$, $B=0$, and using (\ref{1.12}), viz. $(\tau^2+1)Q^{-3/2}_{-1/2+i \tau}(\chi)=\coth\chi Q^{-1/2}_{-1/2+i \tau}(\chi)-Q^{1/2}_{-1/2+i \tau}(\chi)$, then  following a contour integration we obtain
\begin{align}
&\int_{-\infty}^{\infty} d\tau_R Q^{-3/2}_{-1/2+i ( \tau_R-i\epsilon)}(\chi)Q^{-1/2}_{-1/2-i  (\tau_R+i\epsilon)}(\chi)
\nonumber\\
&=\int_{-\infty}^{\infty} d\tau_R\left[\frac{1}{1+(\tau_R-i\epsilon)^2}\left(\frac{\cosh\chi}{\sinh\chi} Q^{-1/2}_{-1/2+i (\tau_R-i\epsilon)}(\chi)-Q^{1/2}_{-1/2+i (\tau_R-i\epsilon)}(\chi)\right)\right]Q^{-1/2}_{-1/2-i  (\tau_R+i\epsilon)}(\chi)
\nonumber\\
&=\int_{-\infty}^{\infty} d\tau_R\left[-\frac{\pi\cosh\chi}{2\sinh^2\chi}\frac{e^{-2\chi\epsilon}}{[1+(\tau_R-i\epsilon)^2](\tau^2_R+\epsilon^2)}-\frac{i\pi}{2\sinh\chi}\frac{e^{-2\chi\epsilon}}{[1+(\tau_R-i\epsilon)^2][\tau_R+i\epsilon]}\right]
\nonumber\\
&=-\frac{\pi^2\cosh\chi e^{-2\chi\epsilon}}{2\epsilon \sinh^2\chi}+\frac{\pi^2\cosh\chi e^{-2\chi\epsilon}}{2\sinh^2\chi }-\frac{\pi^2 e^{-2\chi\epsilon}}{2\sinh\chi},
\label{3.19}
\end{align}
together with
\begin{align}
-i\pi \left(\frac{2\pi}{\sinh\chi}\right)^{1/2}\frac{\Gamma   (2)}{\Gamma   (2)\Gamma   (1)}Q^{-3/2}_{-1/2+2\epsilon}(\chi)=-\frac{\pi^2\cosh\chi e^{-2\chi\epsilon}}{2\epsilon \sinh^2\chi }-\frac{\pi^2e^{-2\chi\epsilon}}{\sinh\chi}.
\label{3.20}
\end{align}
Thus we find that the dominant singular terms in (\ref{3.18}) agree. 

For integer $A=\ell$ and $B=\ell^{\prime}$ we can obtain finite integrals if we keep at least one of $a$ and $b$ non-zero. Specifically, with the form for  $Q^{-1/2-\ell}_{-1/2+i\tau}(\chi)$ with integer $\ell$ given in (\ref{1.10}), the only pole on the real $\tau$ axis is at $\tau=0$. Thus if we replace $\tau$ by $a-b-2i\epsilon$, then given the normalization factor in (\ref{1.10}) the right-hand side of (\ref{3.15}) would be finite if $a-b\neq 0$, $(a-b)^2+m^2\neq 0$, $m=1,...,\ell$. Then the integral over real $\tau_R$ on the left-hand side of (\ref{3.15}) would be finite too. While we could  use the Gram-Schmidt procedure to construct an orthogonal basis for the $Q^{-1/2-\ell}_{-1/2+i ( \tau-i\epsilon)}(\chi)$ functions with zero $a$ and $b$, it would not be an orthonormal one since the integral in (\ref{3.18})  is not finite when $A=\ell$, $B=\ell^{\prime}$. However, as long $a-b\neq 0$, $(a-b)^2+m^2\neq 0$, $m=1,...,\ell$, for the $Q^{-1/2-\ell}_{-1/2+i ( \tau+a-i\epsilon)}(\chi)$, $Q^{-1/2-\ell}_{-1/2-i ( \tau+b+i\epsilon)}(\chi)$ functions we could construct an orthonormal basis. (A simple choice would be $a=\ell$, $b=\ell^{\prime}$.)
 
\section{Contour Integration Procedure}
\label{S4}

\subsection{Integrating the sinc functions}

To integrate the sinc functions that have  appeared above  we use the residue theorem. For complex $\tau$ we have $e^{i \chi\tau}=e^{i \chi \tau_R}e^{-\chi \tau_I}$. For $\chi>0$, the exponential factor $e^{i \chi\tau}$ diverges on the lower half-circle (LHC) of the complex $\tau$  plane while it vanishes on the upper half-circle (UHC) of the complex $\tau$ plane. For $\chi<0$ the behavior is reversed. For $\chi>0$, the exponential factor $e^{-i \chi\tau}$ vanishes on the LHC of the complex $\tau$  plane while it diverges on the UHC of the complex $\tau$ plane. For $\chi<0$ the behavior is reversed. We can decompose the sinc function into $e^{\pm i  \chi\tau}$ terms with one vanishing on the UHC and the other vanishing on the LHC. This will allow us to close the integration contour for each term. However, we are starting from an integral along the real $\tau$ axis and these terms have poles at $\tau=0$. To avoid these poles, we decompose $\sin(\chi\tau)$ into $e^{\pm i  \chi\tau}$ terms and displace the poles above the real $\tau$ axis according to
\begin{align} 
\label{4.1}
\int_{-\infty}^{\infty}\frac{\sin(\chi\tau)}{\tau}d\tau=\frac{ 1}{2i}\int_{-\infty}^{\infty}\frac{e^{i \chi\tau}}{\tau-i \epsilon}d\tau-\frac{1 }{2i}\int_{-\infty}^{\infty}\frac{e^{-i \chi\tau}}{\tau-i \epsilon}d\tau.
\end{align} 
For $\chi>0$ we close the $e^{i\chi\tau}$ contour on the  UHC, and the $e^{-i\chi\tau}$ contour on the  LHC, while for $\chi<0$ we close the $e^{i\chi\tau}$ contour on the  LHC, and the $e^{-i\chi\tau}$ contour on the  UHC. This yields
\begin{align} 
\theta(\chi)\int_{-\infty}^{\infty}\frac{\sin(\chi\tau)}{\tau}d\tau=\pi,\qquad
\theta(-\chi)\int_{-\infty}^{\infty}\frac{\sin( \chi\tau)}{\tau}d\tau=-\pi.
\label{4.2}
\end{align} 
Alternatively, we could  displace the poles below the real $\tau$ axis. This yields
\begin{align} 
\label{4.3}
\int_{-\infty}^{\infty}\frac{\sin(\chi\tau)}{\tau}d\tau=\frac{1 }{2i}\int_{-\infty}^{\infty}\frac{e^{i \chi\tau}}{\tau+i \epsilon}d\tau-\frac{1}{2i}\int_{-\infty}^{\infty}\frac{e^{-i \chi\tau}}{\tau+i \epsilon}d\tau,
\end{align} 
\begin{align} 
\theta(\chi)\int_{-\infty}^{\infty}\frac{\sin(\chi\tau)}{\tau}d\tau=\pi,\qquad  \theta(-\chi)\int_{-\infty}^{\infty}\frac{\sin(\chi\tau)}{\tau}d\tau=-\pi.
\label{4.4}
\end{align} 
Thus no matter on which side of the real $\tau$ axis we put the poles as long as they are all on the same side we obtain 
\begin{align} 
\int_{-\infty}^{\infty}\frac{\sin(\chi\tau)}{\tau}d\tau=\epsilon(\chi)\pi,
\label{4.5}
\end{align} 
where $\epsilon(\chi)=\theta(\chi)-\theta(-\chi)$.

As given in (\ref{2.3}) the  $P^{-1/2-\ell}_{-1/2+i\tau}(\chi)$ functions can be expressed as sums of sinc functions: 
\begin{align} 
\label{4.6}
P^{-1/2-\ell}_{-1/2+i \tau}(\chi)=2\sum_{n=-\ell}^{\ell}\mathcal{R}^{\ell}_{n}(\chi)\frac{\sin(\chi(\tau-i  n))}{\tau-i  n}.
\end{align} 
Recalling that $\mathcal{R}^{\ell}_{n}(\chi)=\mathcal{R}^{\ell}_{-n}(\chi)$ where  $\mathcal{R}^{\ell}_{n}(\chi)$ is given in (\ref{2.5}), then with $\chi>0$  
using contour integration with respect to $\tau$ we obtain  
\begin{align}
\int_{-\infty}^{\infty}d\tau P^{-1/2-\ell}_{-1/2+i \tau}(\chi)=2\pi \sum_{n=-\ell}^{\ell}\mathcal{R}^{\ell}_{n}(\chi),
\label{4.7}
\end{align}
for a single $P^{-1/2-\ell}_{-1/2+i\tau}(\chi)$ function.

\subsection{Integrating products of two sinc functions}

The integral of the product of two $P^{-1/2-\ell}_{-1/2+i\tau}(\chi)$ functions would initially appear to involve a double sum over two sets of poles. However, this is not the case. To be specific, consider the integral of the product of two sinc functions with $\chi_a>\chi_b> 0$, where the product is of the form 
\begin{align} 
\label{4.8}
\frac{\sin(\chi_a\tau)}{\tau}\frac{\sin(\chi_b\tau)}{\tau}=-\frac{1}{4\tau^{2}}{\bigg (}e^{i  (\chi_a+\chi_b)\tau}-e^{i  (\chi_a-\chi_b)\tau}-e^{-i  (\chi_a+\chi_b)\tau}+e^{-i  (\chi_a-\chi_b)\tau}{\bigg )}.
\end{align} 
As constructed, the last two terms in (\ref{4.8}) vanish on the LHC, while the first two terms vanish on the UHC. However, rather than then performing a contour integration, we note that because $\chi_a>\chi_b$ we do not actually need to decompose the $\sin(\chi_b\tau)$ term at all.  Rather, we set 
\begin{align} 
\label{4.9}
\frac{\sin(\chi_a\tau)}{\tau}\frac{\sin(\chi_b\tau)}{\tau}=\frac{1 }{2i}{\bigg (}\frac{e^{i  \chi_a\tau}}{\tau}-\frac{e^{-i  \chi_a\tau}}{\tau}{\bigg )}\frac{\sin(\chi_b\tau)}{\tau}.
\end{align} 
In (\ref{4.9})  the suppression of $e^{-i \chi_a\tau}$ on the LHC dominates over the divergence of $e^{i \chi_b\tau}$ on the LHC, so the factor of $e^{-i \chi_a\tau}$ on its own forces the entire product to vanish on the LHC. Similarly, the suppression $e^{i \chi_{a}\tau}$ on the UHC forces the entire term of which it is a part  to vanish on the UHC. With $\chi_a>\chi_b>0$ we will call the $\sin(\chi_a\tau)$ function the dominant function and refer to its $\chi$ factor as the dominant $\chi$. In the contour integration method we have presented, we only need to decompose the dominant function. The non-dominant sinc function can remain intact and is simply evaluated at the poles of the dominant sinc. Thus as per (\ref{4.1}) to (\ref{4.5}) with all the poles on the same side of the real $\tau$ axis  this yields 
\begin{align} 
\label{4.10}
\int_{-\infty}^{\infty}\frac{\sin(\chi^{}_a\tau)}{\tau}\frac{\sin(\chi^{}_b\tau)}{\tau}d\tau=\pi \chi^{}_b, \ \ \ \ \ \ \ {[}\chi^{}_a>\chi^{}_b{]}.
\end{align} 
Since this same analysis would cause the integral in (\ref{4.10}) to be equal to $\pi\chi_a$ if $\chi_b>\chi_a$, (\ref{4.10}) will thus hold if $\chi_a=\chi_b$. (When $\chi_a=\chi_b$  the necessary asymptotic  suppression for the $e^{\pm i(\chi_a-\chi_b)\tau}/\tau^2$ terms  in (\ref{4.8})  is provided by the $1/\tau^{2}$ factor.)

We can immediately apply this method to integrate the product of two $P^{-1/2-\ell}_{-1/2+i\tau}(\chi)$ functions with respect to $\tau$, to obtain
\begin{align} 
\label{4.11}
&\int_{-\infty}^{\infty}P^{-1/2-\ell}_{-1/2+i \tau}(\chi)P^{-1/2-\ell^{\prime}}_{-1/2+i \tau}(\chi^{\prime})d\tau=2\pi\sum_{n=-\ell}^{\ell} P^{-1/2-\ell^{\prime}}_{-1/2+n}(\chi^{\prime})\mathcal{R}^{\ell}_{n}(\chi), \ \ \ \ {[}\chi\geq\chi^{\prime}{]},
\nonumber\\
&\int_{-\infty}^{\infty}P^{-1/2-\ell}_{-1/2+i \tau}(\chi)P^{-1/2-\ell^{\prime}}_{-1/2+i \tau}(\chi)d\tau=2\pi\sum_{n=-\ell}^{\ell} P^{-1/2-\ell^{\prime}}_{-1/2+n}(\chi)\mathcal{R}^{\ell}_{n}(\chi), \ \ \ \ {[}\chi=\chi^{\prime}{]}.
\end{align} 
As a check we note that when $\ell=0$, $\ell^{\prime}=0$,  and $\chi=\chi^{\prime}$ (\ref{4.11}) immediately recovers (\ref{2.17}).
Thus in general we  decompose the sinc function with the larger value of $\chi$ into exponential functions and then evaluate the other sinc function at the poles of the function with larger $\chi$. If $\chi=\chi^{\prime}$ we can choose which function to decompose. This freedom would allow us to choose to decompose the function with the smaller value of $\ell$ in order to minimize the number of  terms we would have to evaluate.

\subsection{Integrating multiple products of sinc functions and the Borwein integral}

We can generalize the contour integration method presented  to integrals of products of $N+1$ sinc functions. In general, we might have to decompose into exponentials multiple sinc functions in the product, which involves many poles and greatly increases the computational work. However, if the largest $\chi$ (to be labelled $\chi_0$) is greater than the sum of all the other frequencies, then it dominates the entire product, so we then only need to decompose the dominant function (labelled by $n=0$) according to
\begin{align} 
\label{4.12}
\prod^{N}_{n= 0}\frac{\sin(\chi_{n}\tau)}{\tau}=\frac{1 }{2i}{\bigg (}\frac{e^{i \chi\tau}}{\tau-i\epsilon}-\frac{e^{-i \chi\tau}}{\tau-i\epsilon}{\bigg )}\prod^M_{n= 1}\frac{\sin(\chi_{n}\tau)}{\tau}.
\end{align} 
With all $\chi_n$ positive, closing the $e^{i\chi \tau}$ contour on the UHC while closing the $e^{i\chi \tau}$ contour on the LHC yields
\begin{align} 
\label{4.13}
\int_{-\infty}^{\infty}\prod^N_{n= 0}\frac{\sin(\chi_{n}\tau)}{\tau}d\tau=\pi\prod^N_{n= 1}\chi_{n}, \ \ \ \ \ {\bigg [}\chi_{0} \geq\sum^N_{n= 1}\chi_{n}{\bigg ]}.
\end{align}  
This set of integrals  that appear in (\ref{4.13})  are known as Borwein Integrals \cite{Borwein2001}. The standard method for evaluating Borwein integrals makes use of the fact that sinc functions are bandwidth limited. We describe the structure of  Borwein integrals in detail in Appendix A, and in Appendix B and \cite{footnote9} we discuss some general properties of arbitrary bandwidth limited signals from the perspective of contour integration. Some alternate discussion of the role of contour integration in elucidating properties of Borwein integrals may be found in \cite{Labora2025}, a paper that appeared at the same time as \cite{footnote9}.

In addition we can consider putting a polynomial factor $p(\tau)$ with leading order term $\tau^{\nu}$ inside the integral. A polynomial is not a bandwidth limited system, it is not normalizable, and has  a very inconvenient  Fourier transform ($\int dp p^{\nu}e^{ipx} \sim (d/d x)^{\nu}\delta(x)$). The inclusion of a polynomial factor can potentially vitiate  the  determination of  a relevant  integral that is obtained by first Fourier transforming the individual factors of the integrand. In contrast, the contour integration method makes handling the inclusion of a polynomial factor trivial, since we can continue $\tau^{\nu}$ into the complex $\tau$ plane, and the exponential suppression on the circle at infinity that we have been using is not affected by power behavior. Thus all we need to do is evaluate the polynomial terms at the locations of the relevant sinc function poles. As long as $p(\tau)$ has no zeroes at the positions of the relevant sinc function poles, then for the standard Borwein integral for instance this yields
\begin{align} 
\int_{-\infty}^{\infty}\prod_{n=0}^{N}p(\tau)\frac{\sin(\chi_{n}\tau)}{\tau}d\tau=\pi p(0)\prod_{n=1}^{N}\chi_{n}, \ \ \ \ \ {\bigg [}\chi^{}_{0}\geq\sum_{n=1}^{N}\chi_{n}, \ \ N>\mu{\bigg ]},
\label{4.14}
\end{align}  
where $N+1$  is the total number of sinc functions and $\mu$ is the leading order of the polynomial. Similarly, for normalized $P^{-1/2-\ell}_{-1/2+i \tau}(\chi)$ we have 
\begin{align} 
\label{4.15}
\frac{2}{\pi} \int_{-\infty}^{\infty}\frac{1}{\tau^{2} N^{2}_{\ell}(\tau)}{\big [}P^{-1/2-\ell}_{-1/2+i \tau}(\chi){\big ]}^{2}d\tau=2\pi(\ell!)^{2}\mathcal{R}^{\ell}_{0}(\chi)P^{-1/2-\ell}_{-1/2+0}(\chi).
\end{align} 
We have included the additional factor of $1/\tau^{2}$  in this example so that  the integral would converge \cite{footnote10}. For this integral we write one of the $P^{-1/2-\ell}_{-1/2+i \tau}(\chi)$ in terms of  sinc functions as per (\ref{2.3}), and then evaluate the rest of the integral as per the sinc function poles given in (\ref{4.1}) to (\ref{4.5}). However, the factor of $1/\tau^{2}N^{2}_{\ell}(\tau)$ cancels all poles from the decomposing of the sinc functions in  $P^{-1/2-\ell}_{-1/2+i \tau}(\chi)$  except for the pole at $\tau=0$, where $1/\tau^{2}N^{2}_{\ell}(\tau)$  becomes $(\pi(\ell !)^{2}/2$. This example reveals the power of the contour integration method. Not only can the contour integration method handle the inclusion of additional factors $\tau^{\nu}$ inside the integral (such as those in $1/\tau^{2}N^{2}_{\ell}(\tau)$), the inclusion of those factors could potentially simplify the contour integration procedure by canceling out poles and thus reducing the number of terms that need to be calculated in the final sum. 

\subsection{Integrating multiple products of sinc functions and the Nyquist-Shannon sampling theorem}

We can also consider  a product of $M$ sinc functions in which no individual function has a dominant phase over all the others. In this situation rather than decomposing multiple sinc functions, we can instead introduce a dummy  factor of $1=\sin(\chi^{}_{0}\tau)/\sin(\chi^{}_{0}\tau)$ with a phase $\chi_0$  that is greater than or equal to the sum of all the $M$  sinc function $\chi_m$, as  labeled from $m=1$ to $m=M$. We can then decompose the $\sin(\chi^{}_{0}\tau)$ factor according to 
\begin{align} 
\label{4.16}
\frac{\sin(\chi^{}_{0}\tau)}{\sin(\chi^{}_{0}\tau)}\prod^M_{m= 1}\frac{\sin(\chi_{m}\tau)}{\tau}=\frac{1}{2i\sin(\chi^{}_{0}\tau)}{\bigg (}e^{i \chi^{}_{0}\tau}-e^{-i \chi^{}_{0}\tau}{\bigg )}\prod^M_{m= 1}\frac{\sin(\chi_{m}\tau)}{\tau},\ \ \ \ \ {\bigg [}\chi^{}_{0}\geq\sum^M_{m=1}\chi^{}_{m}{\bigg ]},
\end{align} 
in order to identify the  terms for which we can close contours with a suppressed circle at infinity contribution, something we can do even when $\chi^{}_{0}=\sum^M_{m=1}\chi^{}_{m}$, since the circle at  infinity suppression is  then provided by a  $1/\tau^M$ factor. 
The factor of $e^{i \chi^{}_{0}\tau}/\sin(\chi^{}_{0}\tau)$ has poles at $\tau^{}_{n}=n\pi/\chi^{}_{0}$ with integer $n$ and residue $1/\chi^{}_{0}$, viz.
\begin{align}
\label{4.17}
 \tau=\frac{n\pi}{\chi_{0}}+\epsilon: \ \ \frac{e^{i \chi_{0}\tau}}{\sin(\chi_{0}\tau)}\approx \frac{1}{\chi_{0}\epsilon}.
 \end{align} 
By contour integration the integral of a product of sinc functions becomes the discrete sum
\begin{align} 
\label{4.18}
\int_{-\infty}^{\infty}\prod^M_{m= 1}\frac{\sin(\chi_{m}\tau)}{\tau}d\tau=\frac{\pi}{\chi_0}\sum_{n}\prod^M_{m= 1}\frac{\sin(\chi_m n\pi/\chi_0)}{n\pi/\chi_0},
\end{align} 
With $\sin(\chi\tau)/\tau=(1/2)\int_{-\chi}^{\chi}e^{i\omega\tau}d\omega$, the sinc function is a bandwidth limited signal. Moreover, not only is a single sinc function bandwidth limited, as we show in Appendix B, so are products of them such as the product exhibited in (\ref{4.16}). We thus recognize (\ref{4.18}) 
as a specific realization of the Nyquist–Shannon sampling theorem \cite{Nyquist2024,Nyquist2028,Shannon1949} for bandwidth limited signals, with the sampling being done by evaluating (\ref{4.16}) at the $\tau_n=n\pi/\chi$ poles for all integer $n$, a discrete complete basis of sampling points.   We identify $\pi/\chi^{}_{0}=\Delta\tau$ as the spacing between individual  poles. A key feature of the theorem is that  the right-hand side of (\ref{4.18}) is independent of the particular choice of $\chi_0$ as long as it obeys $\chi^{}_{0}\geq\sum^M_{m=1}\chi^{}_{m}$.

The standard proof of the Nyquist–Shannon sampling theorem involves first Fourier transforming a bandwidth limited signal to a space where it is bounded within a finite domain and then using a discrete inverse Fourier transform to recover the function in real space as a discrete sum.  However, that method of deriving the sampling theorem is not well suited to include a polynomial factor such as  $p(\tau)$ in the integrand because polynomials have derivatives of delta function  Fourier transforms. In contrast, the contour integration procedure presented here makes the inclusion of the polynomial factor very easy  to handle since any power of $\tau$ cannot destroy the suppression on the circle at infinity in the complex $\tau$ plane. Consequently, the polynomial just adds to the residue in the contour integration and leads to 
\begin{align} 
\int_{-\infty}^{\infty}p(\tau)\prod_{m=1}^{M}\frac{\sin(\chi^{}_{m}\tau)}{\tau}d\tau=\Delta\tau\sum_{n}p{\big (}n\Delta\tau{\big )}\prod_{m= 1}^{M}\frac{\sin(\chi^{}_{m}n\Delta \tau)}{n\Delta\tau}, \quad {\bigg [}\Delta\tau=\frac{\pi}{\chi^{}_{0}}, \ \ \chi^{}_{0}\geq\sum_{m=1}^{M}\chi^{}_{m}{\bigg ]},
\label{4.19}
\end{align} 
provided  that when the polynomial is included, the integral still  converges.  

The contour method derivation of the Nyquist-Shannon theorem given here is in the context of sinc functions. In Appendix B we present a derivation of the theorem for an arbitrary bandwidth limited function, a derivation that will also readily enable us to incorporate polynomials. Also we note that we could generalize the treatment given in (\ref{4.16}) to any $f(\chi_0\tau)/f(\chi_0\tau)$ that can be decomposed into sums of exponential functions of $\tau$ or sums of exponentials each one multiplied by its own polynomial in $\tau$.

\subsection{Application of the contour method to cosmological fluctuations}

We can apply the contour integration technique to a product of four conical functions of the first kind, a product that appears in  the illustrative $k<0$ conformal gravity model example discussed in  \cite{Mannheim2020,Amarasinghe2021b}. In the model the  cosmic microwave background scalar component of the fluctuation temperature-temperature angular correlation  anisotropy at last scattering is related to 
\begin{align}
I(\ell, K)= \int_{0}^{\infty}d\tau\frac{P(\tau)}{N^{2}_{\ell}(\tau)}P^{-1/2-\ell}_{-1/2+i\tau}(\chi^{}_L)P^{-1/2-K}_{-1/2+i\tau}(\rho_{L})P^{-1/2-\ell}_{-1/2+i\tau}(\chi^{}_{L})P^{-1/2-K}_{-1/2+i\tau}(\rho^{}_{ L}),
\label{4.20}
\end{align}
with $N_{\ell}(\tau)$ being given in (\ref{1.11}).
Here $\rho_L$ is the temporal coordinate and $\chi_L$ is the spatial coordinate for the last photons that scatter against the baryons in the universe  (referred to as last scattering) before the photons stream freely, and typically $P(\tau)=A\tau^{N-4}e^{-\beta^2\tau^2}$, where $A$, $N$ and $\beta$ are real constants \cite{footnote11}.  (The more standard $k=0$, $\Lambda CDM$ model can be discussed analogously, though there the wave functions are Bessel functions.)

With $P^{-1/2-K}_{-1/2+i\tau}(\rho)\rightarrow \tau^{-K-1}$, $P^{-1/2-\ell}_{-1/2+i\tau}(\chi)\rightarrow \tau^{-\ell-1}$, $1/N_{\ell}(\tau)\rightarrow \tau^{\ell+1}$ as $\tau \rightarrow \infty$, then when $\beta=0$ $I(\ell,K)$ will exist if $N-2K-5<0$. While, $I(\ell,K)$ would exist without any such constraint when $\beta$ is non-zero, the $e^{-\beta^2\tau^2}$ term diverges on sections of both the UHC and LHC in the complex $\tau$ plane (explicitly for $\tau=re^{i\theta}$ the divergent regions are $\pi/4 \leq \theta \leq 3\pi/4$, $5\pi/4 \leq \theta \leq 7\pi/4$). Consequently, we cannot apply the contour method to $I(\ell,K)$ when $\beta \neq 0$, so here we shall apply the contour method to the $\beta=0$ case. However we have developed an alternative approach based on properties of bandwidth limited signals that can incorporate $\beta \neq 0$, and will describe it in Appendix B.

With $\beta=0$  we use the contour integration method to evaluate the  temperature-temperature angular correlation integral with $\chi_0=2\chi^{}_L+2\rho^{}_L$ and discrete spacing $\Delta\tau=\pi/(2\chi^{}_L+2\rho^{}_L)$. Thus leads to 
\begin{align}
 &I(\ell.K)= 2\Delta\tau\sum_{n=0}^{\infty}\frac{A(n\Delta\tau)^{N-4}}{N^{2}_{\ell}(n\Delta\tau)}P^{-1/2-\ell}_{-1/2+i  n\Delta\tau}(\chi_L)P^{-1/2-K}_{-1/2+i  n\Delta\tau}(\rho_L)P^{-1/2-\ell}_{-1/2+i  n\Delta\tau}(\chi_L)P^{-1/2-K}_{-1/2+i  n\Delta\tau}(\rho_L).
 \label{4.21}
\end{align}  
The contour integration technique thus enables us to replace an infinite integral by a potentially more tractable discrete sum, since for large $n\Delta\tau$ we have 
\begin{align}
\frac{A(n\Delta\tau)^{N-4}}{N^{2}_{\ell}(n\Delta\tau)}P^{-1/2-\ell}_{-1/2+i  n\Delta\tau}(\chi_L)P^{-1/2-K}_{-1/2+i  n\Delta\tau}(\rho_L)P^{-1/2-\ell}_{-1/2+i  n\Delta\tau}(\chi_L)P^{-1/2-K}_{-1/2+i  n\Delta\tau}(\rho_L)\rightarrow (n\Delta\tau)^{N-2K-6},
\label{4.22}
\end{align}
a vanishingly small form at large $n$ since we have already imposed $N-2K-5<0$. With this example we thus demonstrate the power of the interplay of sinc functions  and the contour method that we have presented in this paper. 
\appendix
\numberwithin{equation}{section}
\setcounter{equation}{0}
\section{The Borwein Integral}

The Borwein integral is a striking and non-intuitive phenomenon discovered by Borwein and Borwein in integrals of progressions of products of sinc functions.
It takes the form  \cite{Borwein2001}
\begin{align}
 \int_{0}^{\infty}\frac{\sin(x/1)}{x/1}{\it dx}&=\frac{\pi}{2},
 \nonumber\\
\int_{0}^{\infty}\frac{\sin(x/1)}{x/1}\frac{\sin(x/3)}{x/3}{\it dx}&=\frac{\pi}{2},
\nonumber\\ 
\int_{0}^{\infty}\frac{\sin(x/1)}{x/1}\frac{\sin(x/3)}{x/3}\frac{\sin(x/5)}{x/5}{\it dx}&=\frac{\pi}{2},
\nonumber\\ 
\int_{0}^{\infty}\frac{\sin(x/1)}{x/1}\frac{\sin(x/3)}{x/3}\frac{\sin(x/5)}{x/5}\frac{\sin(x/7)}{x/7}... \frac{\sin(x/13)}{x/13}{\it dx}&=\frac{\pi}{2},
\label{A.1}
\end{align}
all of which integrals evaluate to the same  $\pi/2$ before the pattern suddenly breaks at $\tfrac{\sin(x/15)}{x/15}$ where
\begin{align}
\int_{0}^{\infty}\frac{\sin(x/1)}{x/1}\frac{\sin(x/3)}{x/3}\frac{\sin(x/5)}{x/5}\frac{\sin(x/7)}{x/7}... \frac{\sin(x/15)}{x/15}{\it dx}= \frac{\pi}{2}\left(1-\frac{6879714958723010531}{935615849440640907310521750000}\right).
\label{A.2}
\end{align}
The leading sinc function in the sequence has a frequency of one and the break in the pattern occurs when the sum of the sub-leading frequencies becomes greater than one, i.e., according to 
\begin{align}
 \frac{1}{3}+\frac{1}{5}+\frac{1}{7}+\frac{1}{9}+\frac{1}{11}+\frac{1}{13}=.955 < 1,
\nonumber\\  \frac{1}{3}+\frac{1}{5}+\frac{1}{7}+\frac{1}{9}+\frac{1}{11}+\frac{1}{13}+\frac{1}{15}=1.022 > 1.
\label{A.3}
\end{align}
Furthermore, the amount by which the value of the integral in (\ref{A.2}) departs from  $\pi/2$ was found to be expressible in terms of the sub-leading frequencies according to
\begin{align}
\frac{6879714958723010531}{935615849440640907310521750000}=\frac{(1/3+1/5+1/7+1/9+1/11+1/13+1/15-1)^{7}}{2^{6}\cdot 7! \cdot 1/3 \cdot 1/5 \cdot 1/7 \cdot 1/9 \cdot 1/11 \cdot 1/13 \cdot 1/15}.
\label{A.4}
\end{align}

The first explanation of this behavior was given by Borwein and Borwein and made use of the Fourier transform of the sinc function. The sinc function is simply the finite range  integral of  a cosine:
\begin{align}
\frac{\sin x}{x}=\int_{0}^{1}\cos(\omega x){\it d\omega}=\frac{1}{2}\int_{0}^{1}[e^{i\omega x}+e^{-i\omega x}]d\omega,
\label{A.5}
\end{align}
and hence its Fourier transform is just a sum of phases over a finite domain, viz. a bandwidth limited functionj By taking the Fourier transform of all the sinc functions, Borwein and Borwein turned the integral into a sequence of repeated convolutions of square waves. The repeated convolutions smear out the wave packet in frequency space, and then the resulting wave packet is evaluated at its center point. Borwein and Borwein showed through this method that when the sum of the sub-leading frequencies is less than the leading frequency the center point of the Fourier transformed wave packet remains at a value of one after repeated convolutions but when the sum of the sub-leading frequencies becomes greater than the leading frequency the value of the center point drops to less than one after repeated convolutions. From this insight, Borwein and Borwein found a more general form of the integral. For a product of sinc functions of the form $\sin(\nu_{n}x)/\nu_{n}x$ with positive frequencies $\nu_{n}\geq\nu_{n+1}$, Borwein and Borwein obtained 
\begin{align}
\int_{-\infty}^{\infty}\prod^N_{n= 0}\frac{\sin(\chi_{n}\tau)}{\tau}d\tau=\pi\prod^N_{n= 1}\chi_{n}, \ \ \ \ \ {\bigg [}\chi_{0} \geq\sum^N_{n= 1}\chi_{n}{\bigg ]}.
\label{A.6}
\end{align}
We recognize (\ref{A.6}) as (\ref{4.13}), as  derived in this paper via contour integration, an approach that, as noted above, allows for the inclusion of non-sinc functions such as polynomials.  

In regard to why there is a break at $1/15$ in (\ref{A.3}), we note that this is the value at which the largest frequency  in the (\ref{A.1}) type products  is no longer greater than the sum of all the other frequency values combined. Consequently, the contour integration no longer converges on the circle at infinity if we only decompose the sinc function with the largest frequency value. In order to make the contour integration converge on the circle at infinity in this case, we have to decompose more than just one sinc function, which introduces additional poles that  had not  previously been involved.

\section{The Nyquist-Shannon sampling theorem via a bandwidth limited signal analysis}

Consider the function $B(\tau)$ defined by the finite domain integral 
\begin{align} 
B(\tau)=\frac{1}{2\pi}\int_{-\chi}^{\chi}B_{\omega}e^{i\omega\tau} d\omega,
\label{B.1}
\end{align}
where $B_{\omega}$ is some function of $\omega$ and the only dependence on $\tau$ is in the $e^{i\omega\tau}$ phase. $B(\tau)$ is a bandwidth limited signal with maximum frequency $\omega=\pm\chi$ where $\chi$ is positive. Specifically, the Fourier transform of $B(\tau)$ is defined as
\begin{align} 
\tilde{B}(\omega)=\int_{-\infty}^{\infty}B(\tau)e^{-i\omega\tau}d\tau,\qquad B(\tau)=\frac{1}{2\pi}\int_{-\infty}^{\infty}\tilde{B}(\omega)e^{i\omega\tau}d\omega,
\label{B.2}
\end{align}
with it being assumed that $B(\tau)$ vanishes sufficiently fast as $\tau \rightarrow \pm \infty$ so that $\tilde{B}(\omega)$ exists. 
Inserting (\ref{B.1}) into (\ref{B.2})  gives
\begin{align} 
\tilde{B}(\omega)= \frac{1}{2\pi}\int_{-\infty}^{\infty}d\tau\int_{-\chi}^{\chi}d\omega^{\prime}B_{\omega^{\prime}}e^{i\omega^{\prime}\tau}e^{-i\omega\tau},
\label{B.3}
\end{align}
so that 
\begin{align} 
\tilde{B}(\omega)=\int_{-\chi}^{\chi}B_{\omega^{\prime}}\delta(\omega^{\prime}-\omega) d\omega^{\prime}=\theta(\chi-|\omega|)B_{\omega}.
\label{B.4}
\end{align}
With the Fourier transform of $B(\tau)$ thus being zero outside the range $-\chi<\omega<\chi$, finite domain integrals such as the one associated with $B(\tau)$ are thus bandwidth limited in Fourier space. A specific example of a bandwidth limited signal is given by the integral representation of $P^{-1/2-K}_{-1/2+i\tau}(\chi)$  that is exhibited in  (\ref{1.6}). In this section we shall examine some general consequences of arbitrary bandwidth limited signals.

To begin we note that the product of two  bandwidth limited signals is itself a bandwidth limited signal. To be  specific, consider the functions 
\begin{align} 
B^1(\tau)=\int_{-\chi_{1}}^{\chi_{1}}B^1_{\omega}e^{i\omega\tau}d\omega, \qquad B^2(\tau)=\int_{-\chi_{2}}^{\chi_{2}}B^2_{\omega^{\prime}}e^{i\omega^{\prime}\tau}d\omega^{\prime}.
\label{B.5}
\end{align}
On taking their product we set $\omega_+=\omega+\omega^{\prime}$,  $\omega_-=\omega-\omega^{\prime}$, so that 
the frequency modes combine as $e^{i\omega\tau}e^{i\omega^{\prime}\tau}=e^{i\omega_+\tau}$, where $-\chi_{1}-\chi_{2}\leq \omega_+\leq \chi_{1}+\chi_{2}$. 
On integrating over $\omega_-$  we find that the product is a bandwidth limited signal of the form 
\begin{align} 
B^1(\tau)B^2(\tau)=\int_{-\chi_{1}-\chi_{2}}^{\chi_{1}+\chi_{2}}C_{\omega_+}e^{i\omega_+\tau}d\omega_+,
\label{B.6}
\end{align}
where  $C_{\omega_+}$ is an appropriate function of $\omega_+$.  $B^1(\tau)B^2(\tau)$ is thus a bandwidth limited signal whose Fourier transform vanishes for all $|\omega_+|>\chi_{1}+\chi_{2}$.

If we now introduce a third bandwidth limited signal $B^{3}(\tau)$, then $B^1(\tau)B^2(\tau)B^3(\tau)$ is a product of two bandwidth limited signals (viz. $B^1(\tau)B^2(\tau)$ and $B^3(\tau)$) to thus also be a bandwidth limited signal, one whose Fourier transform vanishes for   $|\omega_+|>\chi_{1}+\chi_2+\chi_3$.
More generally, as well as any  multi-bandwidth product $B^1(\tau)B^2(\tau)....B^n(\tau)$ any function defined by a finite multi-dimensional integral of the form
\begin{align} 
B(\tau)=\int_{-\chi_{1}}^{\chi_{1}}d\omega_{1}\int_{-\chi_{2}}^{\chi_{2}}d\omega_{2}...\int_{-\chi_{n}}^{\chi_{n}}d\omega_{n}B(\omega_{1},\omega_{2},... ,\omega_{n})e^{i(\omega_{1}+\omega_{2}+...+\omega_{n})\tau}
\label{B.7}
\end{align}
is a bandwidth limited signal  of the form
\begin{align} 
B(\tau)=\int_{-\chi_{1}-\chi_{2}-... -\chi_{n}}^{\chi_{1}+\chi_{2}+...+\chi_{n}}C_{\omega_+}e^{i\omega_+\tau}d\omega_+,
\label{B.8}
\end{align}
with appropriate function $C_{\omega_+}$, where $\omega_+=\omega_{1}+\omega_{2}+...+\omega_{n}$, and where $-\chi_{1}-\chi_{2}-... -\chi_{n}\leq\omega_+\leq \chi_{1}+\chi_{2}+...+\chi_{n}$.  

Furthermore, if we multiply a bandwidth limited signal  by a polynomial factor $p(\tau)$ with  power $\tau^N$ where $N$ is integer, then assuming that the Fourier transform of $P(\tau)B(\tau)$ exists, the Fourier transform is then given by 
\begin{align} 
\int_{-\infty}^{\infty}P(\tau)B(\tau)e^{-i\omega\tau}d\tau&=\int_{-\infty}^{\infty}\tau^Ne^{-i\omega\tau}d\tau\frac{1}{2\pi}\int_{-\chi}^{\chi}B_{\omega^{\prime}}e^{i\omega^{\prime}\tau}d\omega^{\prime}=i^N\frac{d^N}{d\omega^N}\left[\int_{-\infty}^{\infty}e^{-i\omega\tau}d\tau\frac{1}{2\pi}\int_{-\chi}^{\chi}B_{\omega^{\prime}}e^{i\omega^{\prime}\tau}d\omega^{\prime}\right]
\nonumber\\
&=i^N\frac{d^N}{d\omega^N}\left[\int_{-\chi}^{\chi}B_{\omega^{\prime}}\delta(\omega^{\prime}-\omega)d\omega^{\prime}\right]
=i^N\frac{d^N}{d\omega^N}\left[B_{\omega}\theta (\chi>|\omega|)\right].
\label{B.9}
\end{align}
Thus even though polynomials such as $P(\tau)$ are not themselves bandwidth limited, nonetheless, once $B(\tau)$ is bandwidth limited, then so is the $P(\tau)B(\tau)$ product.  

\subsection{Evaluating integrals of products of bandwidth limited functions and non-polynomial-bounded functions}

If we want to evaluate a potentially non-analytically-tractable   integral of the form $\int d\tau A(\tau)B(\tau)$, viz.  the product of some bandwidth limited function $B(\tau)$ with a non-bandwidth-limited function $A(\tau)$ that does not converge  on either the entire upper or entire lower half circle at infinity in the complex $\tau$ plane (such as $e^{-\beta^2\tau^2}$ for instance), we cannot use the above polynomial procedure. However, there is something else we can do instead. Specifically, we can Fourier transform $B(\tau)$ and use its bandwidth limited structure to evaluate $\tilde{B}(\omega)$ as a contour integral in the complex $\tau$ plane instead. Then we can Fourier  transform $\tilde{B}(\omega)$  back and evaluate $\int d\tau A(\tau)B(\tau)$ with the resulting $B(\tau)$, a procedure that can potentially give us a tractable  $\int d\tau A(\tau)B(\tau)$. To be explicit, we note that with the definition of $\tilde{B}(\omega)$ for a bandwidth limited signal, viz. 
\begin{align} 
\tilde{B}(\omega)=\theta(\chi-|\omega |)\int_{-\infty}^{\infty}d\tau B(\tau)e^{-i\omega\tau}=\frac{1}{2\pi}\theta(\chi-|\omega|)\int_{-\infty}^{\infty}d\tau\int_{-\chi}^{\chi}d\omega^{\prime}B_{\omega^{\prime}}e^{i\omega^{\prime}\tau}e^{-i\omega\tau},
\label{B.10}
\end{align}
the leading frequency modes in $\tilde{B}(\omega)$ are $e^{\pm 2i\chi\tau}$ since both $\omega$ and $\omega^{\prime}$ are restricted to the domain  $(-\chi,\chi)$. We introduce a factor of $1=\sin(\bar{\chi}\tau)/\sin(\bar{\chi}\tau)$ where $\bar{\chi}=\chi+\delta$ with $\delta$ small and positive, to give
\begin{align} 
\tilde{B}(\omega)&= \frac{1}{2\pi}\theta(\chi-|\omega|)\int_{-\infty}^{\infty}d\tau\int_{-\chi}^{\chi}d\omega^{\prime}\frac{(e^{i\bar{\chi}\tau}-e^{-i\bar{\chi}\tau})}{2i\sin(\bar{\chi}\tau)}B_{\omega^{\prime}}e^{i\omega^{\prime}\tau}e^{-i\omega\tau}.
\label{B.11}
\end{align}
With $\bar{\chi}$ positive, for large positive $\tau_I$ we have 
\begin{align} 
 \frac{e^{i\bar{\chi}\tau}}{2i\sin(\bar{\chi}\tau)}&\rightarrow -e^{-2\tau_I(\bar{\chi}+\delta)},\qquad 
- \frac{e^{-i\bar{\chi}\tau}}{2i\sin(\bar{\chi}\tau)}\rightarrow 1,
\label{B.12}
\end{align}
while for large negative $\tau_I$ we have
\begin{align} 
 \frac{e^{i\bar{\chi}\tau}}{2i\sin(\bar{\chi}\tau)}&\rightarrow 1,\qquad 
- \frac{e^{-i\bar{\chi}\tau}}{2i\sin(\bar{\chi}\tau)}\rightarrow -e^{2\tau_I(\bar{\chi}+\delta)}.
\label{B.13}
\end{align}
Since the leading values of $e^{i(\omega^{\prime}-\omega)\tau}$, viz. $e^{\pm 2i\chi\tau}$, are of the form $e^{\mp 2\chi\tau_I}$ when $\tau$ is pure imaginary, the presence of the $\delta$ factor ensures that the portion of the integrand in (\ref{B.11}) that contains the $e^{i\bar{\chi}\tau}$ factor converges on the UHC in the complex $\tau$ plane, while the portion of the integrand in (\ref{B.11}) that contains the $-e^{-i\bar{\chi}\tau}$ factor converges on the LHC.

The $1/\sin(\bar{\chi}\tau)$ factor has an infinite series of poles at $\tau=n\pi/\bar{\chi}^2+i\epsilon=n\pi/\chi-n\pi\delta/\chi^2+i\epsilon$ for all integer $n$, poles that for convenience we have situated above the real $\tau$ axis in the complex $\tau$ plane. On closing the contour these poles only contribute to  the term with the $e^{i\chi \tau}$ factor and have residues 
\begin{align} 
{\rm Res}{\bigg [}\frac{e^{i\bar{\chi}\tau}}{2i\sin(\bar{\chi}\tau)}B(\tau)e^{-i\omega\tau},\tau=\frac{n\pi}{\bar{\chi}}+i\epsilon{\bigg ]}=\frac{1}{2i\bar{\chi}}B(n\pi/\bar{\chi})e^{-i\omega n\pi/\bar{\chi}},
\label{B.14}
\end{align}
with contour integration thus giving 
\begin{align} 
\tilde{B}(\omega)=\theta(\chi-|\omega|)\frac{\pi}{\chi}\sum_{n=-\infty}^{\infty}B(n\pi/\chi)e^{-i\omega n\pi/\chi},
\label{B.15}
\end{align}
after setting $\delta=0$.
Thus we can now  expand the original bandwidth limited  function $B(\tau)$  as
\begin{align} 
B(\tau)=\frac{1}{2\pi}\frac{\pi}{\chi}\sum_{n=-\infty}^{\infty}B(n\pi/\chi)\int_{-\chi}^{\chi}e^{i\omega(\tau-n\pi/\chi)}d\omega=
\frac{1}{\chi}\sum_{n=-\infty}^{\infty}B(n\pi/\chi)\int_{0}^{\chi}\cos(\omega(\tau-n\pi/\chi))d\omega.
\label{B.16}
\end{align}
And if $B(\tau)$ is symmetric we obtain 
\begin{align} 
B(\tau)&=\frac{1}{\chi}\sum_{n=-\infty}^{\infty}B(n\pi/\chi)\int_{0}^{\chi}\cos(\omega\tau)\cos(\omega n\pi/\chi)d\omega
\nonumber\\
&=\frac{1}{\chi}\sum_{n=-\infty}^{\infty}B(n\pi/\chi)\frac{(\tau\sin(\chi\tau) \cos(n\pi)-(n\pi/\chi)\cos(\chi\tau)\sin(n\pi)}{(\tau^2-n^2\pi^2/\chi^2)}=\sum_{n=-\infty}^{\infty}B(n\pi/\chi)\frac{(-1)^n\chi\tau\sin(\chi\tau)}{(\chi^2\tau^2-n^2\pi^2)}
\nonumber\\
&=\frac{1}{2}\sum_{n=-\infty}^{\infty}B(n\pi/\chi)\left(\frac{\sin(\chi\tau-n\pi)}{(\chi\tau-n\pi)}+\frac{\sin(\chi\tau+n\pi)}{(\chi\tau+n\pi)}\right).
\label{B.17}
\end{align}
We recognize (\ref{B.17}) as being built out  of sinc functions, just like  (\ref{4.18}).  With (\ref{B.17}) we have a new derivation of the Nyquist-Shannon sampling theorem for a general bandwidth limited signal. While our introduction of a  $1=\sin(\bar{\chi}\tau)/\sin(\bar{\chi}\tau)$  factor in (\ref{B.11}) might seem innocuous, its  $\sin(\bar{\chi}\tau)$ denominator produces poles at $\tau=n\pi/\chi$ for all integer $n$, to thus provide an appropriate complete basis of $n\pi/\chi$ modes  with which to do the sampling.  Moreover, since $1=\sin(\bar{\chi}\tau)/\sin(\bar{\chi}\tau)$  is an exact relation, the sampling generated by it is exact also.

Our contour integration  method is particularly convenient for computing the Fourier transform of a polynomial $\it p(\tau)$ multiplied by a bandwidth limited signal $B(\tau)$. The polynomial does not affect the behavior of $B(\tau)$ on the circle at infinity in the complex $\tau$ plane, so it is just evaluated at the poles of the factor of $1/\sin(\chi\tau)$, and gives
\begin{align} 
p(\tau)B(\tau)=\frac{1}{2\pi}\frac{\pi}{\chi}\sum_{n=-\infty}^{\infty}p(n\pi/\chi)B(n\pi/\chi)\int_{-\chi}^{\chi}e^{i\omega(\tau-n\pi/\chi)}d\omega.
\label{B.18}
\end{align}
For a non-polynomial-bounded function $A(\tau)$ we can still use (\ref{B.17}) as we now show.

\subsection{Application of the bandwidth limited signal procedure to cosmological fluctuations}

For the tensor fluctuation component contribution to the cosmic microwave background anisotropy  in the   $k<0$ cosmological model discussed in \cite{Mannheim2025},  the relevant integrated Sachs-Wolfe effect integral is of the form
\begin{align} 
C_{\ell}&=2\pi\int _{\rho_L}^{\rho_0}d\rho_1\int_{\rho_L}^{\rho_0}d\rho_2\int_0^{\infty} d\tau \frac{\partial E^*(\rho_1,\tau)}{\partial \rho_1}\frac{\partial E(\rho_2,\tau)}{\partial \rho_2}P(\tau)E^*_{\tau,\ell}(\chi_1)E_{\tau,\ell}(\chi_2)\bigg{|}_{\chi=\rho_0-\rho}.
\label{B.19}
\end{align}
Here $C_{\ell}$ is the probability for a fluctuation of angular momentum $\ell$, with $\ell\geq 2$ since the modes are tensor.  
$\rho$ is the temporal coordinate with current era value  $\rho_0$ and last scattering value $\rho_L$.   $\chi=\rho_0-\rho$ is the spatial coordinate with current era value zero and and last scattering value $\chi_L=\rho_0-\rho_L$. And with a closed form expression for $ P^{-3/2}_{-1/2+i\tau}(\rho)$ being given in   (\ref{1.7}) and with $N_{\ell}(\tau)$ being given in (\ref{1.11})\, the spatial and temporal mode functions are given  by
\begin{align}
E_{\tau,\ell}(\chi)&=(-1)^{\ell+1}\left(\frac{2}{\pi}\right)^{1/2}\frac{1}{N_{\ell}(\tau)}\frac{P^{-1/2-\ell}_{-1/2+i\tau}(\chi)}{\sinh^{5/2}\chi},
\nonumber\\
E(\rho,\tau)&=\sinh^{3/2}\rho P^{-3/2}_{-1/2+i\tau}(\rho)=\frac{1}{\tau(\tau^2+1)}\left(\frac{2}{\pi}\right)^{1/2}\left[\cosh\rho\sin(\tau\rho)-\tau \sinh\rho\cos(\tau\rho)\right],
\nonumber\\
\partial_{\rho}E(\rho,\tau)&=\left(\frac{2}{\pi}\right)^{1/2}\sinh\rho\frac{\sin(\tau\rho)}{\tau},
\label{B.20}
\end{align}
in a cosmological model with negative 3-curvature and a cosmological constant.  The function $P(\tau)$ has a typical form $P(\tau)=A\tau^{N-4}e^{-\beta^2\tau^2}$, where $A$, $N$, and $\beta$ are real constants. It is the presence of the $e^{-\beta^2\tau^2}$ term  that prevents convergence on the circle at infinity in the complex $\tau$ plane.  However, we can apply (\ref{B.17}) to the bandwidth limited part of (\ref{B.19}) (a part that does not include the $e^{-\beta^2\tau^2}$ term),   viz. 
\begin{align} 
B(\tau)=\frac{1}{N^2_{\ell}(\tau)}{\bigg [}\int_{0}^{\chi_{L}}d\chi\sinh(\rho_{0}-\chi)\frac{\sin(\tau(\rho_{0}-\chi))}{\tau}\frac{P^{-1/2-\ell}_{-1/2+i\tau}(\chi)}{\sinh^{5/2}\chi}{\bigg ]}^{2},
\label{B.21}
\end{align}
because as per (\ref{1.6}) $P^{-1/2-\ell}_{-1/2+i\tau}(\chi)$ is bandwidth limited   with frequency limited to the range $(-\chi,\chi)$, $P^{-3/2}_{-1/2+i\tau}(\rho )$ is bandwidth limited  with frequency limited to $-(\rho_{0}-\chi)\leq \omega \leq \rho_0-\chi$, the product of multiple  bandwidth limited signals is a bandwidth limited signal, and the product of a bandwidth limited signal  with a polynomial factor like $1/N^2_{\ell}(\tau)$ is also a bandwidth limited signal \cite{footnote12}. Since $1/N_{\ell}(\tau)$ asymptotically behaves as $\tau^{\ell+1}$, the $P^{-1/2-\ell}_{-1/2+i\tau}(\chi)$ each have a leading factor of $\tau^{-\ell-1}$, and the $\sin((\rho_{0}-\chi)\tau)/\tau$ term provides  for an additional factor of $1/\tau$, the needed convergence of $B(\tau)\sim 1/\tau^2$ at large $\tau$ is met. With these various components of $B(\tau)$ being bandwidth limited, the allowed frequency range for them is $(-\chi_{d},\chi_d)$, and with $\chi$ and $\rho$ having opposite signs (a negative $\rho$ being measured forward from the big bang (i.e., getting less negative) and a positive $\chi$ being measured backwards from $\chi=0$ today),  the $\chi_d$ appropriate for the integrated 
Sachs-Wolfe $(0, \chi_L)$ range is given by  
\begin{align} 
\chi_{d}=2|\chi_{L}-\rho_{0}|+2\chi_{L}.
\label{B.22}
\end{align}

We introduce a factor of $\sin(\bar{\chi}_{d}\tau)/\sin(\bar{\chi}_{d}\tau)$ where $\bar{\chi}_d=\chi_d+\delta$. Following the procedure described earlier  we compute the Fourier transform
\begin{align} 
\tilde{B}(\omega)=\int_{-\infty}^{\infty}d\tau\frac{\sin(\bar{\chi}_{d}\tau)}{\sin(\bar{\chi}_{d}\tau)}\frac{1}{N^2_{\ell}(\tau)}\bigg{[}\int_{0}^{\chi_{L}}d\chi \sinh(\rho_{0}-\chi)\frac{\sin((\rho_{0}-\chi)\tau)}{\tau}\frac{P^{-1/2-\ell}_{-1/2+i\tau}(\chi)}{\sinh^{5/2}\chi}\bigg {]}^{2}e^{-i\omega\tau}.
\label{B.23}
\end{align} 
As per (\ref{B.17}) and (\ref{B.18}) this gives  
\begin{align} 
B(\tau)&=\frac{1}{\chi_{d}}\sum_{n=-\infty}^{\infty}\frac{1}{N^2_{\ell}(n\pi/\chi_d)}\bigg{[}\int_{0}^{\chi_{L}}d\chi \sinh(\rho_{0}-\chi)\frac{\sin((\rho_{0}-\chi)n\pi/\chi_d)}{n\pi/\chi_d}\frac{P^{-1/2-\ell}_{-1/2+i n\pi/\chi_d}(\chi)}{\sinh^{5/2}\chi}\bigg{]}^{2}
\nonumber\\ 
&\times \int_{0}^{\chi_{d}}\cos(\omega\tau)\cos(\omega n\pi/\chi_d)d\omega
\nonumber\\
&=\frac{1}{\chi_{d}}\sum_{n=-\infty}^{\infty}\frac{1}{N^2_{\ell}(n\pi/\chi_d)}\bigg{[}\int_{0}^{\chi_{L}}d\chi \sinh(\rho_{0}-\chi)\frac{\sin((\rho_{0}-\chi)n\pi/\chi_d)}{n\pi/\chi_d}\frac{P^{-1/2-\ell}_{-1/2+i n\pi/\chi_d}(\chi)}{\sinh^{5/2}\chi}\bigg{]}^{2}
\nonumber\\
&\times\frac{1}{2} \int_{-\chi_d}^{\chi_{d}}e^{i\omega\tau}cos(\omega n\pi/\chi_d)d\omega
\label{B.24}
\end{align}
Inspection of (\ref{B.24}) shows that  $B(\tau)$ is a symmetric function of $\tau$, with the continuous parameter $\tau$ that appears in (\ref{B.23}) having been replaced by a discrete sum over the points $\tau=n\pi/\chi_d$, so that  the only remaining dependence on $\tau$ is in the $\cos(\omega\tau)$ or $e^{i\omega\tau}$  factors.  $B(\tau)$ is thus in the form given in (\ref{B.1}), to thus not only have components that are bandwidth limited, but to actually be a bandwidth limited function itself.  We thus show that there is a quite intricate connection between cosmological perturbation theory and the Nyquist-Shannon sampling theorem.

Given that the only dependence on $\tau$ in (\ref{B.24}) is in the $\cos(\omega\tau)$ factor, we can now include the $e^{-\beta^2\tau^2}$ term and  compute the ensuing  $\tau$  integral in $C_{\ell}$ analytically. This gives
\begin{align} 
\int_{-\infty}^{\infty}\tau^{N-4}e^{-\beta^2\tau^2}\cos(\omega\tau)d\tau=\frac{1}{2}\beta^{(3-N)}\Gamma((N-3)/2)\Phi((N-3)/2,1/2,-\omega^2/4\beta^2),
\label{B.25}
\end{align}
where $\Phi(\alpha,\gamma;z)$ is the confluent hypergeometric function $_1F_1(\alpha;\gamma;z)=1+\alpha z/
\gamma+\alpha(\alpha+1)z^2 /\gamma(\gamma+1)2!+....$.
Lastly, we integrate over $\omega$, and obtain
\begin{align} 
C_{\ell}&=\frac{4A}{\pi^2\chi_{d}}\sum_{n=-\infty}^{\infty}\frac{1}{N^2_{\ell}(n\pi/\chi_d)}  {\bigg [}\int_{0}^{\chi_{L}}d\chi \sinh(\rho_{0}-\chi)\frac{\sin((\rho_{0}-\chi)n\pi/\chi_d)}{n\pi/\chi_d}\frac{P^{-1/2-\ell}_{-1/2-i n\pi/\chi_d}(\chi)}{\sinh^{5/2}\chi}{\bigg ]}^{2}
\nonumber\\ 
&\times \frac{1}{2}\int_{0}^{\chi_{d}}\beta^{(3-N)}\Gamma((N-3)/2)\Phi((N-3)/2,1/2,-\omega^{2}/4\beta^{2})\cos(\omega n\pi/\chi_d)d\omega.
\label{B.26}
\end{align}
Thus through use of  the treatment of the Nyquist-Shannon sampling theorem that we have developed here, we have been able to replace the integration  on $\tau$ in (\ref{B.19}) by the discrete sum on $n$ given in (\ref{B.26}).  While (\ref{B.26}) involves an infinite sum on $n$, for large $n$ $C_{\ell}$ behaves as $1/n^2$, which is very beneficial for numerically evaluating $C_{\ell}$. We provide an actual application of  (\ref{B.26}) to  cosmology  in \cite{Mannheim2025}. In fact, it was the our need to treat the cosmology described in \cite{Mannheim2025} that occasioned our study of  bandwidth limited functions  in the first place.


\begin{thebibliography}{99}
\bibitem{footnote1} The standard associated Legendre polynomials $P^m_{\ell}(z)$ have $\nu=\ell$ and $\mu=m$.
\bibitem{footnote2} The second form for $N_{\ell}(\tau)$ given in (\ref{1.11}) can be shown to be equal to the first form by using $\tau^2+\ell^2=(\ell+i\tau)(\ell-i\tau)$, $\Gamma(\ell+i\tau)=(\ell-1+i\tau)\Gamma(\ell-1+i\tau)$, and working down until reaching $\Gamma(1-i\tau)\Gamma(i\tau)$, which is equal to $\pi/\sin(i\pi\tau)$.
\bibitem{footnote5} To check for $P^{-5/2}_{-1/2+i\tau}(\chi)$ we use (\ref{1.12}) to construct it from the forms for  $P^{-3/2}_{-1/2+i\tau}(\chi)$  and  $P^{-1/2}_{-1/2+i\tau}(\chi)$  given  in (\ref{1.7}).
\bibitem{footnote6} While there is also  $1/\cos(K\pi)$ factor, which  could give rise to  poles when $K$ is half-integer (a pole that would be $\tau$-independent), at such half-integer $K$ the factor $P^{-1/2-K}_{-1/2+i\tau}(\chi)-\Gamma(i\tau-K)P^{1/2+K}_{-1/2+i\tau}(\chi)/\Gamma(i\tau+1+K)$ vanishes, with (\ref{3.6}) giving a full accounting of the pole structure.
\bibitem{footnote12} While (\ref{B.21}) involves an overall  finite  integration range $(0,\chi_L)$, in and of itself this does not make the full $B(\tau)$ a bandwidth limited signal, since its overall $\tau$ dependence is not simply an $e^{i\omega\tau}$ phase. That the full $B(\tau)$ is a bandwidth limited signal is because it is composed of functions that themselves are bandwidth limited. Thus as in (\ref{B.8}) the allowed bandwidth limited range is fixed by the  bandwidth limited functions contained in the full $B(\tau)$, and is exhibited in (\ref{B.22}).
\end{thebibliography}
\end{document}